\documentclass[journal]{IEEEtran}
\ifCLASSINFOpdf \else \fi

\usepackage{graphicx}
\usepackage[cmex10]{amsmath}
\usepackage{subfigure}
\usepackage{amssymb}
\usepackage{bbm}

\usepackage[table]{xcolor}
\usepackage{tabularx}
\usepackage{multirow}
\usepackage{acro}[=v3] 
 \DeclareAcronym{6G}{short = 6G, long  = sixth-generation ,tag = abbrev}
\DeclareAcronym{5G}{short = 5G, long  = fifth-generation ,tag = abbrev}
\DeclareAcronym{4G}{short = 4G, long  = fourth-generation ,tag = abbrev}
\DeclareAcronym{ISAC}{short = ISAC , long  = integrated sensing and communication,tag = abbrev}
\DeclareAcronym{MMSE}{short = MMSE , long  = minimum mean squared error ,tag = abbrev}
\DeclareAcronym{RMSE}{short = RMSE , long  = root mean squared error ,tag = abbrev}
\DeclareAcronym{DFT}{short = DFT, long  = discrete Fourier transform ,tag = abbrev}
\DeclareAcronym{IDFT}{short = IDFT, long  = inverse discrete Fourier transform ,tag = abbrev}
\DeclareAcronym{FFT}{short = FFT, long  = fast Fourier transform ,tag = abbrev}
\DeclareAcronym{ISFFT}{short = ISFFT, long  = inverse symplectic finite Fourier transform ,tag = abbrev}
\DeclareAcronym{IFFT}{short = IFFT, long  = inverse fast Fourier transform ,tag = abbrev}
\DeclareAcronym{SFFT}{short = SFFT, long  = symplectic finite Fourier transform ,tag = abbrev}
\DeclareAcronym{V2X}{short = V2X, long  = vehicle-to-everything ,tag = abbrev}
\DeclareAcronym{NTN}{short = NTN, long  = non-terrestrial networks ,tag = abbrev}
\DeclareAcronym{MSE}{short = MSE, long  = mean square error ,tag = abbrev}
\DeclareAcronym{CSI}{short = CSI , long  = channel state information ,tag = abbrev}
\DeclareAcronym{MIMO}{short = MIMO, long  = multiple-input multiple-output ,tag = abbrev}
\DeclareAcronym{BW}{short = Bw, long  = bandwidth ,tag = abbrev}
\DeclareAcronym{BS}{short = BS, long  = base station ,tag = abbrev}
\DeclareAcronym{LS}{short = LS, long  = least square ,tag = abbrev}
\DeclareAcronym{SNR}{short = SNR, long  = signal-to-noise ratio ,tag = abbrev}
\DeclareAcronym{NMSE}{short = NMSE, long  = normalized mean square error ,tag = abbrev}
\DeclareAcronym{PAPR}{short = PAPR, long  = peak to average power ratio ,tag = abbrev}
\DeclareAcronym{ISI}{short = ISI, long  = inter-symbol interference ,tag = abbrev}
\DeclareAcronym{ICI}{short = ICI, long  = inter-carrier interference ,tag = abbrev}
\DeclareAcronym{CP}{short = CP, long  = cyclic prefix ,tag = abbrev}
\DeclareAcronym{CPP}{short = CPP, long  = chirp-periodic prefix ,tag = abbrev}
\DeclareAcronym{ZP}{short = ZP, long  = zero padding ,tag = abbrev}
\DeclareAcronym{ZS}{short = ZS, long  = zero suffix ,tag = abbrev}
\DeclareAcronym{ZF}{short = ZF, long  = zero forcing ,tag = abbrev}
\DeclareAcronym{RZP}{short = RZP, long  = reduced-zero padded ,tag = abbrev}
\DeclareAcronym{BER}{short = BER, long  = bit error rate ,tag = abbrev}
\DeclareAcronym{SIC}{short = SIC, long  = successive interference cancellation ,tag = abbrev}
\DeclareAcronym{AWGN}{short = AWGN, long  = additive white Gaussian noise,tag = abbrev}
\DeclareAcronym{SINR}{short = SINR, long  = signal-to-interference-plus-noise ratio ,tag = abbrev}\DeclareAcronym{SIR}{short = SIR, long  = signal-to-interference ratio ,tag = abbrev}
\DeclareAcronym{BPSK}{short = BPSK, long  = binary phase shift keying ,tag = abbrev}
\DeclareAcronym{QPSK}{short = QPSK, long  = quadrature phase shift keying ,tag = abbrev}
\DeclareAcronym{PDF}{short = PDF, long  = probability density function ,tag = abbrev}
\DeclareAcronym{CDF}{short = CDF, long  = cumulative distribution function,tag = abbrev}
\DeclareAcronym{1D}{short = 1D, long  = 1-dimensional ,tag = abbrev}
\DeclareAcronym{2D}{short = 2D, long  = 2-dimensional ,tag = abbrev}
\DeclareAcronym{3D}{short = 3D, long  = three-dimensional ,tag = abbrev}
\DeclareAcronym{QAM}{short = QAM, long  = quadrature amplitude modulation ,tag = abbrev}
\DeclareAcronym{CE}{short = CE, long  = channel estimation ,tag = abbrev}
\DeclareAcronym{FMCW}{short = FMCW, long  = frequency modulated continuous wave ,tag = abbrev}
\DeclareAcronym{JSAC}{short = JSAC, long  = joint sensing and communication ,tag = abbrev}
\DeclareAcronym{OTFS}{short = OTFS, long  = orthogonal time-frequency space ,tag = abbrev}
\DeclareAcronym{SE}{short = SE, long  =spectral efficiency,tag = abbrev}
\DeclareAcronym{MP}{short = MP, long  = message passing ,tag = abbrev}
\DeclareAcronym{OMP}{short = OMP, long  = orthogonal matching pursuit ,tag = abbrev}
\DeclareAcronym{ML}{short = ML, long  = maximum likelihood ,tag = abbrev}
\DeclareAcronym{OOK}{short = OOK, long  = on-off keying ,tag = abbrev}
\DeclareAcronym{OFDM}{short = OFDM, long  = orthogonal frequency division multiplexing ,tag = abbrev}
\DeclareAcronym{AFDM}{short = AFDM, long  = affine frequency division multiplexing ,tag = abbrev}
\DeclareAcronym{OCDM}{short = OCDM, long  = orthogonal chirp division multiplexing ,tag = abbrev}
\DeclareAcronym{ODDM}{short = ODDM, long  = orthogonal delay-Doppler division multiplexing,tag = abbrev}
\DeclareAcronym{DDM}{short = DDM, long  = delay-Doppler map,tag = abbrev}
\DeclareAcronym{FD}{short = FD, long  = full duplex,tag = abbrev}
\DeclareAcronym{PRI}{short = PRI, long  = pulse repetition interval,tag = abbrev}
\DeclareAcronym{LTV}{short = LTV, long  = linear time-varying  ,tag = abbrev}
\DeclareAcronym{RIS}{short = RIS, long  = Reconfigurable
intelligent surfaces, tag = abbrev}
\DeclareAcronym{FrFT}{short = FrFT, long  = fractional Fourier transform  ,tag = abbrev}
\DeclareAcronym{IDAFT}{short = IDAFT, long  = inverse discrete affine Fourier transform,tag = abbrev}
\DeclareAcronym{DAFT}{short = DAFT, long  = discrete affine Fourier transform,tag = abbrev}
\DeclareAcronym{PRS}{short = PRS, long  = positioning reference signal,tag = abbrev}
\DeclareAcronym{AFT}{short = AFT, long  = Affine Fourier transform,tag = abbrev}
\DeclareAcronym{STFT}{short = STFT, long  = short-time Fourier transform,tag = abbrev}
\DeclareAcronym{DD}{short = DD-coupling, long  = Doppler-delay coupling,tag = abbrev}
\DeclareAcronym{SBL}{short = SBL, long  = sparse bayesian learning,tag = abbrev}
\DeclareAcronym{EPA}{short = EPA, long  = embedded pilot-aided,tag = abbrev}
\DeclareAcronym{PCTD}{short = PCTD, long  = post-coded time domain,tag = abbrev}
\DeclareAcronym{MRC}{short = MRC, long  = maximum ratio combiner,tag = abbrev}
\DeclareAcronym{RRC}{short = RRC, long  = root raised cosine,tag = abbrev}
\DeclareAcronym{EVM}{short = EVM, long  = error vector magnitude,tag = abbrev}
\DeclareAcronym{EVA}{short = EVA, long  = extended vehicular A,tag = abbrev}
\DeclareAcronym{SI}{short = SI, long  = self interference,tag = abbrev}
\DeclareAcronym{CSC}{short = CSC, long  = circularly shifted chirp,tag = abbrev}
\DeclareAcronym{LFM}{short = LFM, long  = linear frequency modulated,tag = abbrev}
\DeclareAcronym{LoRa}{short = LoRa, long  = long range,tag = abbrev}
\DeclareAcronym{Tx}{short = Tx, long  = transmitting,tag = abbrev}
\DeclareAcronym{Rx}{short = Rx, long  = receiving,tag = abbrev}
\DeclareAcronym{RCS}{short = RCS, long  = radar cross section,tag = abbrev}
\DeclareAcronym{CFO}{short = CFO, long  = carrier frequency offset,tag = abbrev}
\DeclareAcronym{CPI}{short = CPI, long  = coherent processing interval,tag = abbrev}
\DeclareAcronym{WSS}{short = WSS, long  = wide sense stationary,tag = abbrev}
\DeclareAcronym{IID}{short = i.i.d., long  = independent and identically distributed,tag = abbrev}
\DeclareAcronym{NOMA}{short = NOMA, long  = non orthogonal multiple access,tag = abbrev}
\DeclareAcronym{RDM}{short = RDM, long  = range-Doppler map,tag = abbrev}
\DeclareAcronym{LoS}{short = LoS, long  = line of sight,tag = abbrev}
\DeclareAcronym{NLoS}{short = NLoS, long  = non-line of sight,tag = abbrev}
\DeclareAcronym{IoT}{short = IoT, long  = internet of things,tag = abbrev}
\DeclareAcronym{TF}{short = TF, long  = time-frequency,tag = abbrev}
\DeclareAcronym{3GPP}{short = 3GPP, long  = 3rd Generation Partnership project,tag = abbrev}
\DeclareAcronym{OTSM}{short = OTSM, long  = orthogonal time sequency modulation,tag = abbrev}
\DeclareAcronym{DMRS}{short = DMRS, long  = demodulation reference signal,tag = abbrev}
\DeclareAcronym{PBCH}{short = PBCH, long  = physical broadcast channel,tag = abbrev}
\DeclareAcronym{NR}{short = NR, long  = new radio,tag = abbrev}
\DeclareAcronym{PLS}{short = PLS, long  = physical-layer security,tag = abbrev}
\DeclareAcronym{UE}{short = UE, long  = user-equipment,tag = abbrev}
\DeclareAcronym{LTI}{short = LTI, long  = linear time invariant,tag = abbrev}
\DeclareAcronym{CFR}{short = CFR, long  = channel frequency response,tag = abbrev}
\DeclareAcronym{TDL}{short = TDL, long  = tapped delay line,tag = abbrev}
\DeclareAcronym{CDL}{short = CDL, long  = cluster delay line,tag = abbrev}
\DeclareAcronym{CIR}{short = CIR, long  = channel impulse response,tag = abbrev}
\DeclareAcronym{MPCs}{short = MPCs, long  = multi-path components,tag = abbrev}
\DeclareAcronym{PSD}{short = PSD, long  = power spetcrum density,tag = abbrev}
\DeclareAcronym{PDP}{short = PDP, long  = power delay profile,tag = abbrev}
\DeclareAcronym{PRB}{short = PRB, long  = physical resource block,tag = abbrev}
\DeclareAcronym{WSSUS}{short = WSSUS, long  = wide sense stationary uncorrelated scatter,tag = abbrev}
\DeclareAcronym{SSB}{short = SSB, long  = synchronization signal block,tag = abbrev}
\DeclareAcronym{DPSSB}{short = DP-SSB, long  = dual purpose SSB,tag = abbrev}
\DeclareAcronym{SSS}{short = SSS, long  = secondary synchronization signal,tag = abbrev}
\DeclareAcronym{PSS}{short = PSS, long  = primary synchronization signal,tag = abbrev}
\DeclareAcronym{RE}{short = RE, long  = resource element,tag = abbrev}
\DeclareAcronym{LPSS}{short = LP-SS, long  = low-power synchronization signal,tag = abbrev}
\DeclareAcronym{IA}{short = IA, long  = initial access,tag = abbrev}
\DeclareAcronym{DL}{short = DL, long  = downlink,tag = abbrev}
\DeclareAcronym{TD}{short = TD, long  = time domain,tag = abbrev}
\DeclareAcronym{mMTC}{short = mMTC, long  = massive machine-type communication,tag = abbrev}
\DeclareAcronym{LTE}{short = LTE, long  = long-term evolution,tag = abbrev} 
\DeclareAcronym{NB}{short = NB, long  = narrowband,tag = abbrev}
\DeclareAcronym{LPWUS}{short = LP-WUS, long  = low-power wake-up signal,tag = abbrev}
\DeclareAcronym{PCID}{short = PCID, long  = physical cell identity,tag = abbrev}
\DeclareAcronym{DT}{short = DT, long  = discrete-time,tag = abbrev}
\DeclareAcronym{TO}{short = TO, long  = timing offset,tag = abbrev}
\DeclareAcronym{CM}{short = CM, long  = complex multiplication,tag = abbrev}
\DeclareAcronym{LMMSE}{short = LMMSE, long  = linear minimum mean-square error,tag = abbrev}
\usepackage{algpseudocode}
\usepackage[linesnumbered,ruled,vlined]{algorithm2e}
\usepackage{nomencl}
\usepackage{array}

\usepackage{url}
\usepackage{cite}

\usepackage{xcolor}
\usepackage{etoolbox}

\apptocmd{\thebibliography}{\color{black}}{}{}

\usepackage[compact]{titlesec}
\titlespacing{\section}{1.0pt}{*1.0}{*0}
\titlespacing{\subsection}{1.1pt}{*1.1}{*0}
\titlespacing{\subsubsection}{0.3pt}{*0}{*0}
\begin{document}
\bstctlcite{IEEEexample:BSTcontrol}
\author{
\IEEEauthorblockN{Sadiq Iqbal, Abdelali Arous,~\IEEEmembership{Graduate~Student~Member,~IEEE}, and H\"{u}seyin Arslan,}
\IEEEmembership{Fellow, IEEE}
\thanks{The authors I. Sadiq, A. Arous, and H. Arslan, are with the Department of Electrical and Electronics Engineering, Istanbul Medipol University, Istanbul, 34810, Turkey, (e-mail: sadiq.iqbal@std.medipol.edu.tr; abdelali.arous@std.medipol.edu.tr; huseyinarslan@medipol.edu.tr).}
}

\title{Dual-Purpose SSB Design for Joint Time-Frequency Synchronization Using Relocated DMRS}

\maketitle
\begin{abstract}
The synchronization signal block (SSB) is designed to synchronize a capable user-equipment (UE) with the base-station (BS) in recent cellular networks. However, the same SSB cannot be utilized by a narrowband, low-complexity, and low-power internet of things (IoT) device for synchronization. To address this limitation, this paper proposes a dual-purpose SSB (DP-SSB) to enable low-complexity IoT synchronization using the existing SSB, while ensuring uninterrupted conventional UE operation. Specifically, selected demodulation reference signal (DMRS) symbols are contiguously mapped to some physical resource blocks (PRBs) in the second and fourth symbols of the DP-SSB. These DMRS symbols are further optimized using a receiver-aware least square (LS) approach to impose a distinct time-domain structure at the output of the IoT front-end. At the IoT receiver, a reduced-rate observation of the DP-SSB is obtained after narrowband filtering and decimation that is exploited for energy detection and sequential time-frequency (TF) synchronization. For the UE, primary synchronization signal (PSS) and secondary synchronization signal (SSS)-based synchronization is preserved, while the effect of relocated DMRS on the channel estimation and physical broadcast channel (PBCH) reception is assessed. The performance of the proposed DP-SSB is evaluated in terms of detection probability, TF synchronization accuracy, IoT receiver computational complexity, and UE-side PBCH decoding. Simulation results validate the analytical detection characterization and demonstrate reliable low-power IoT synchronization performance compared with the benchmark synchronization schemes while maintaining conventional UE operation. 
\end{abstract}


\section{Introduction}
\IEEEPARstart{T}{he} transition from \ac{5G} to \ac{6G} wireless networks is expected to support an unprecedented scale of intelligent and interconnected services, enabling emerging applications such as health care, remote education, fully autonomous driving, and smart cities \cite{8766143}. Among these technologies, \ac{IoT} is envisioned as a primary driver, interconnecting billions of devices with future cellular networks \cite{9369324}. Regardless of the application, every cellular device must first establish reliable \ac{IA} and acquire \ac{TF} synchronization with the serving \ac{BS}. However, the stringent energy, coverage, and hardware constraints of low-power \ac{IoT} limit the receiver from processing the wideband \ac{SSB} at the original sampling rate without increasing the front-end and baseband processing, motivating the need for a more efficient low-complexity synchronization solution \cite{9369324, 9200376}. Although numerous \ac{IoT} technologies have been developed, ranging from short-range solutions such as near-field communication, radio frequency identification, Bluetooth, and ZigBee \cite{fi11010016}, to low-power wide-area network technologies such as \ac{LoRa} \cite{9501038}, most are tailored for specific deployment scenarios with allocated and different synchronization mechanisms. For cellular \ac{IoT}, the \ac{3GPP} introduced \ac{NB}-\ac{IoT} to provide extended coverage, low-power consumption, reduced device complexity, and low-cost connectivity \cite{barellini2023time}. Nevertheless, \ac{NB}-\ac{IoT} employs dedicated \ac{NB} synchronization signals for low-power receiver processing, but its synchronization architecture remains distinct from the cellular \ac{UE} synchronization framework.

As cellular networks have evolved, the synchronization frameworks have undergone several architectural changes to accommodate new service requirements \cite{8675913}. \ac{NB}-\ac{IoT} also adopted the \ac{LTE} synchronization philosophy by introducing \ac{NB} synchronization signals and the corresponding cell-search procedures tailored to bandwidth-constrained receivers \cite{7918505}. Several studies have investigated and contributed to improving this synchronization architecture. In \cite{shimura2018physical}, frequency diversity reception has been proposed by transmitting narrow \ac{PSS} and \ac{SSS} over multiple \ac{PRB}s to improve synchronization robustness, while cross-correlation and auto-correlation-based techniques have been developed to jointly improve timing and frequency acquisition while reducing narrow \ac{SSS} detection complexity \cite{kadambar2018low}. Furthermore, in \cite{sun2019enabling}, the authors proposed a frequency-hopping synchronization scheme in which enhanced \ac{NB} synchronization signals are mapped across all fourteen \ac{OFDM} symbols within a subframe and combined with contiguous \ac{TD} repetitions and orthogonal codes to improve synchronization robustness and mitigate false peak detections under challenging channel conditions. Although these approaches improve synchronization reliability and reduce receiver complexity, they continue to rely on dedicated \ac{NB} signaling and remain structurally independent of the \ac{SSB}-based synchronization framework.

In contrast, the \ac{PSS}, \ac{SSS}, \ac{PBCH}, and \ac{DMRS} are integrated into the \ac{SSB} in \ac{5G} \ac{NR}, which provides a fundamental synchronization framework for the conventional \ac{UE}. Significant research efforts have been made to enhance the synchronization capability using the resources available within the \ac{SSB} \cite{8675913,9312170}. In \cite{ohiwa2025physical}, the authors investigated robust \ac{PSS} and \ac{PCID} detection under large \ac{CFO} and severe phase noise by combining \ac{CP}-based autocorrelation with cross-correlation-based frequency offset estimation. Similarly, authors in \cite{li2022pbch} enhanced \ac{PBCH} reception by exploiting the extended \ac{PBCH} repetition structure within the \ac{SSB} to improve synchronization performance under low \ac{SNR}, large frequency offsets, and high-mobility scenarios. Furthermore, in \cite{chong2021efficient}, the known \ac{PBCH}-\ac{DMRS} have also been exploited to further accelerate the cellular synchronization procedure via efficiently detecting \ac{DMRS} sequences. However, these approaches are tailored for wideband conventional \ac{UE}s, where they don't address whether the conventional \ac{SSB} can simultaneously provide a low-complexity synchronization opportunity for low-power \ac{IoT} devices while preserving the synchronization procedure and \ac{PBCH} reception of the \ac{UE}.

Low-power communication for \ac{IoT} devices is further explored as a relevant research direction. In \cite{10453166}, the design of \ac{LPWUS} has been studied to reduce the activity and processing requirements of low-power devices. Moreover, low-\ac{PAPR} multichannel \ac{OOK} \cite{9013137}, \ac{DFT}-spread-\ac{OFDM}-based \ac{OOK} \cite{11048969}, sequence-based \ac{OOK} multiplexed with \ac{OFDM} \cite{sahin2018sequence}, and an \ac{OFDM}-based \ac{OOK} transceiver for wake-up radio and ambient-\ac{IoT} applications \cite{zegrar2026ofdm} have been studied and demonstrate that a reduced-rate observation can substantially lower the processing requirements of a low-power \ac{IoT} receiver. However, these approaches utilize a dedicated low-power signaling framework rather than exploiting the synchronization resources already available in the periodically transmitted \ac{SSB}. From these works, a fundamental mismatch between legacy cellular \ac{IoT} synchronization, conventional \ac{NR} synchronization, and low-power receiver operation is revealed. Existing \ac{IoT} synchronization methods rely upon dedicated \ac{NB} signals, which add signaling overhead and might not be as frequent as conventional \ac{SSB}. Whereas the conventional synchronization framework targets wideband \ac{SSB} reception at the \ac{UE}, low-power signaling techniques generally require an additional waveform to initiate low-power communication. Thus, reusing an existing \ac{SSB} for low-complexity \ac{IoT} synchronization requires a synchronization structure that remains observable after the reduced-rate processing at the \ac{IoT} without disrupting the \ac{UE} synchronization. 

To address this mismatch, this paper proposes a \ac{DPSSB} design in which reallocated and contiguous \ac{LS}-optimized \ac{DMRS} symbols of specific \ac{OFDM} symbols are utilized to enable an \ac{IoT} device to achieve low-complexity synchronization while preserving classical \ac{SSB}-based synchronization for \ac{UE}s. The \ac{DMRS} symbols are optimized to have desirable \ac{TD} correlation properties when filtered around the allocated frequency components, allowing simple detection and synchronization for low-complexity and narrowband receivers, without altering the general frame of the \ac{OFDM}-based \ac{SSB} signaling. The main contributions of this work are summarized as follows:
{\begin{enumerate}
    \item A novel \ac{DPSSB} configuration is proposed in which $K$ selected \ac{DMRS} \ac{RE}s are contiguously placed into a single or multiple \ac{PRB}s of the second and fourth \ac{OFDM} symbols. The remaining \ac{DMRS} and \ac{PBCH} symbols are remapped to uniformly occupy the remaining \ac{RE}s without altering the general structure of the SSB frame. 
    \item A receiver-aware \ac{LS} optimization is derived for the relocated \ac{DMRS} \ac{RE}s to synthesize a sharp-peak shaped \ac{LPSS} in the \ac{TD} after filtering the corresponding spectral components. The \ac{LS} formulation incorporates the known \ac{IoT} receiver filtering and decimation operations together with the deterministic contributions of the remaining \ac{PBCH} and \ac{DMRS} resources to obtain the \ac{LPSS} structure directly in the reduced subspace of the receiver.
    \item Consequently, a low-complexity synchronization framework is developed that exploits the synthesized \ac{LPSS} for energy-based \ac{DPSSB} detection, correlation-based timing acquisition, coarse \ac{CFO} estimation over predefined hypothesis sets, and is followed by fine \ac{CFO} estimation via \ac{FD} phase comparison of the two modified \ac{OFDM} symbols. 
    \item A comprehensive analytical evaluation of the proposed \ac{DPSSB} is performed in terms of detection probability, \ac{TF} synchronization accuracy, frame alignment, and receiver computational complexity. Furthermore, the effect of restrictions imposed by the decimation and filter order on the \ac{TO} and \ac{CFO} is analyzed simultaneously. At the \ac{UE} side, the impact of the modified \ac{DMRS} configuration on the \ac{PBCH} detection is quantified using \ac{LS} and \ac{LMMSE} channel estimation schemes to demonstrate the tradeoff between synchronization and \ac{PBCH} acquisition introduced by the \ac{DPSSB}.
\end{enumerate}

The rest of the paper is organized as follows:  Section II presents the proposed \ac{DPSSB} resource mapping with \ac{LS} optimization incorporating an \ac{IoT} receiver front-end. Section III develops the \ac{DPSSB} detection and the proposed sequential synchronization framework for the narrowband \ac{IoT} device. Following that, performance analysis is carried out in Section IV for both \ac{IoT} and \ac{UE} devices, with simulation results presented in Section V, and the conclusion is drawn in Section VI.

\textit{Notation:} Bold uppercase $\mathbf A$, bold lowercase $\mathbf a$, and unbold letters $A$ and $a$ denote matrices, column vectors, and scalar values, respectively. $\mathbb C^{M\times N}$ denotes the space of $M\times N$ complex-valued matrices. $(\cdot)^{\mathrm H}$ denotes the Hermitian transpose, $\|\cdot\|_2$ is the Euclidean norm, $\odot$ denotes element-wise multiplication, $\Re\{\cdot\}$ returns
the real part, and $\lfloor\cdot\rfloor$ denotes the floor operation. $\dagger$ denotes the Moore–Penrose pseudo-inverse and $\mathbf{F}$ is normalized DFT matrix of size $N$ and entries $e^{-j2\pi kn/N}$. For a subspace $\mathcal{S}\subset\mathbb{C}^{M}$, $\mathcal{S}^{\perp}$ denotes its orthogonal complement.
The symbol $j$ denotes the imaginary unit with $j^2=-1$.

\section{DP-SSB Signal Shaping and Frame Design}
Consider a network consisting of a \ac{BS} serving multiple \ac{UE}s and \ac{IoT} devices distributed throughout the coverage area. Unlike conventional architectures that rely on dedicated gateways or separate \ac{IoT} \ac{BS}, the proposed scheme uses the same \ac{BS} to simultaneously serve both device types through a unified synchronization infrastructure. The following subsections first describe the conventional \ac{SSB}, then introduce the proposed \ac{DPSSB} resource-grid modifications, and finally derive the \ac{LS}-based waveform optimization applied to the relocated \ac{DMRS} symbols.

\subsection{Conventional SSB}
The conventional synchronization framework relies on the \ac{SSB}, consisting of four contiguous \ac{OFDM} symbols carrying $N_c$ subcarriers denoted by the index set $\mathcal{K}_{\mathrm{SSB}}
=\{0,1,\ldots,N_c-1\}$, as depicted in \figurename~\ref{fig_2}(a). The \ac{FD} vector associated with the $\ell$-th \ac{OFDM} symbol is denoted by $\mathbf{x}_{\ell}\in\mathbb{C}^{N_c\times1}$, where
$\ell\in\{0,1,2,3\}$. The first \ac{OFDM} symbol carries the \ac{PSS} and its $k$-th \ac{RE} is expressed as
\begin{equation}
\text{x}_{\mathrm{0}}[\hat{k}]
=\begin{cases}1 - 2c_{\mathrm{PSS}}[k'], 
& \hat{k} = k_{\mathrm{PSS}}^{\min} + k', \\
0, & \text{otherwise},
\end{cases}
\label{eq:pss}
\end{equation}
where $\hat{k}\in\mathcal{K}_{\mathrm{SSB}}$,
$k'\in\{0,\ldots,126\}$,
$c_{\mathrm{PSS}}[k']$ denotes the binary \ac{PSS} sequence, and $k_{\mathrm{PSS}}^{\min}$ its starting subcarrier. The second \ac{OFDM} symbol contains the \ac{PBCH} symbols together with periodically distributed \ac{DMRS} \cite{chong2021efficient}. It is represented by $\mathbf{x}_{1}\in\mathbb{C}^{N_c\times1}$ and the \ac{RE} is expressed as
\begin{equation}
\text{x}_1[\hat{k}]=
\begin{cases}d_{\mathrm{D}}^{(1)}[m],&
\hat{k}=k_{\mathrm{D}}^{(1)}(m),\\d_{\mathrm{P}}^{(1)}[q],&\hat{k}=k_{\mathrm{P}}^{(1)}(q),\\\end{cases}
\label{eq:pbchdmrs}
\end{equation}
where $d_{\mathrm{D}}^{(1)}[m]$ and $d_{\mathrm{P}}^{(1)}[q]$ represent the $m$-th \ac{DMRS} and the $q$-th \ac{PBCH} symbol, respectively. The mappings $k_{\mathrm{D}}^{(1)}(m)$ and
$k_{\mathrm{P}}^{(1)}(q)$ specify their relative subcarrier locations in the second \ac{OFDM} symbol. The third \ac{OFDM} symbol contains the \ac{SSS} together with the \ac{PBCH} and \ac{DMRS} \ac{RE}s, represented by $\mathbf{x}_{2}\in\mathbb{C}^{N_c\times1}$ and is expressed as
\begin{equation}
\text{x}_2[\hat{k}]=
\begin{cases}
1-2c_{\mathrm{SSS}}[k'],&
\hat{k}=k_{\mathrm{SSS}}^{\min}+k',\\
d_{\mathrm D}^{(2)}[m],&
\hat{k}=k_{\mathrm D}^{(2)}(m),\\
d_{\mathrm P}^{(2)}[q],&
\hat{k}=k_{\mathrm P}^{(2)}(q),\\0,&\text{otherwise},
\end{cases}
\label{sss}
\end{equation}
where $c_{\mathrm{SSS}}[k']$ represents the binary \ac{SSS} sequence, $k_{\mathrm{SSS}}^{\min}$ its starting subcarrier, and
$k_{\mathrm D}^{(2)}(m)$ and $k_{\mathrm P}^{(2)}(q)$ denote the \ac{DMRS} and \ac{PBCH} mappings, respectively. The fourth \ac{OFDM} symbol, indicated by 
$\mathbf{x}_{3}\in\mathbb{C}^{N_c\times1}$, contains \ac{PBCH} and \ac{DMRS} symbols and follows the same allocation developed in
\eqref{eq:pbchdmrs}, using the symbol-specific mappings
$k_{\mathrm D}^{(3)}(m)$ and $k_{\mathrm P}^{(3)}(q)$. Overall, these \ac{FD} \ac{OFDM} symbol vectors form the conventional \ac{SSB} resource grid as $\mathbf{X}^{\mathrm{SSB}}=\left[\mathbf{x}_0,
\mathbf {x}_1,\mathbf {x}_2,\mathbf {x}_3
\right]\in\mathbb{C}^{N_c \times 4},$ whose $(\hat{k},\ell)$-th \ac{RE} is denoted by
$X^{\mathrm{SSB}}[\hat{k},\ell]=\text{x}_\ell[\hat{k}]$.
\begin{figure}[!t]
\centering
\includegraphics[width=3.3in]{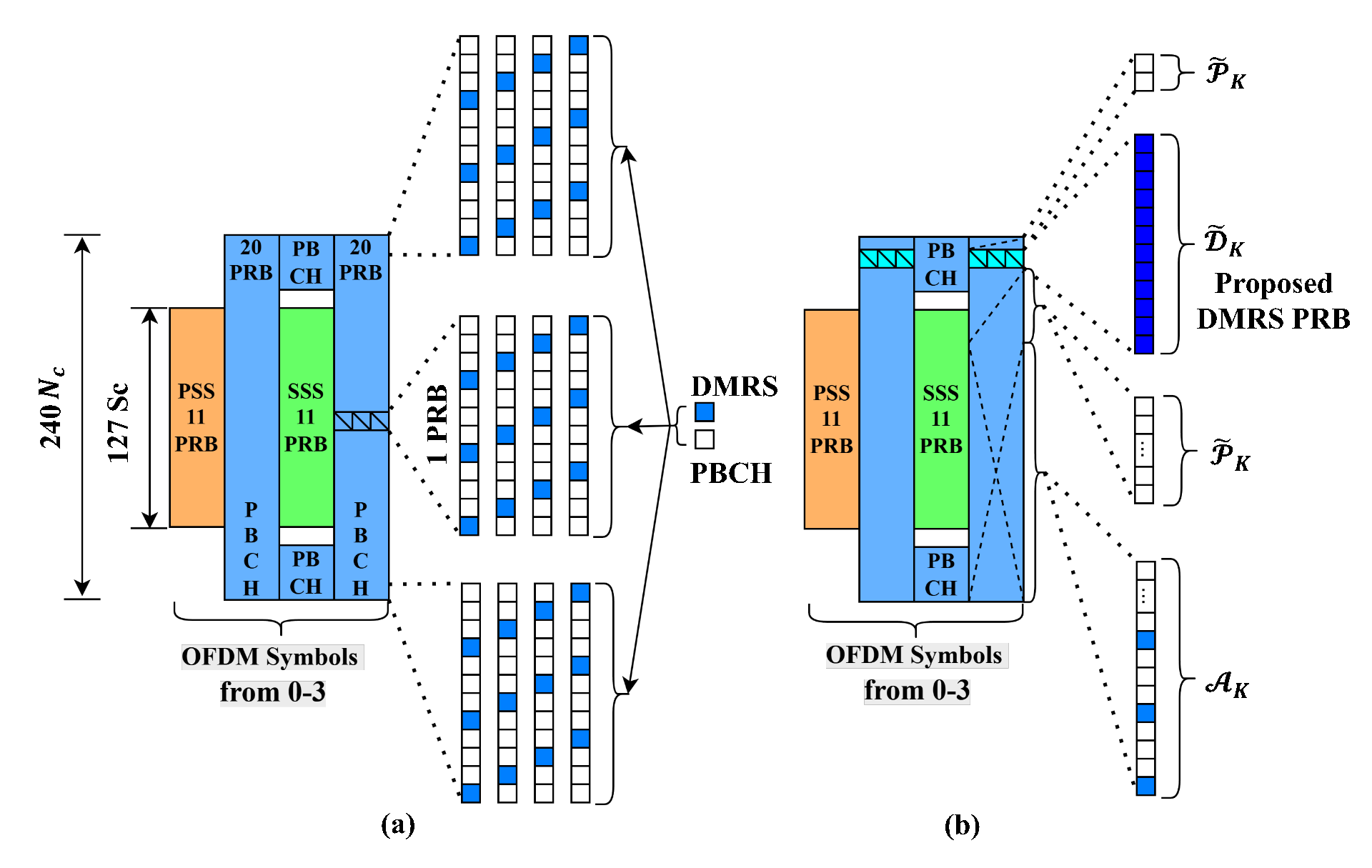}
\caption{\ac{SSB} resource-grid structure: (a) conventional \ac{PBCH}-\ac{DMRS} distribution and (b) proposed \ac{DPSSB} remapping illustrated for $K=12$, where the selected \ac{DMRS} symbols are relocated to the contiguous set $\widetilde{\mathcal{D}}_{K}$ and the \ac{PBCH} symbols occupy $\widetilde{\mathcal{P}}_{K}$.}
\label{fig_2}
\end{figure} 
To fit in the $N$-point \ac{DFT} size ($N>N_c$), a subcarrier-embedding matrix $\mathbf{E}$ is defined that maps the $N_c$ subcarriers onto their corresponding physical $N$-point inverse-\ac{DFT} bins while setting the remaining $N-N_c$ entries to zero, as $\bar{\mathbf X}^{\mathrm{SSB}} =\mathbf E\mathbf X^{\mathrm{SSB}} \in \mathbb C^{N\times4}.$ Therefore, the \ac{SSB} in \ac{DT} is written as
\begin{equation}
s_{\ell}^{\mathrm{SSB}}[n]
=\frac{1}{\sqrt{N}}\sum_{ \hat{k}=0}^{N-1}\bar{\text{X}}^{\mathrm{SSB}}[ \hat{k},\ell]e^{j2\pi \frac{ \hat{k}n}{N}},
\label{eq:ssb_td_cp}
\end{equation}
where $n,k=0,\ldots,N-1$ corresponds to the $N$ \ac{OFDM} samples and total number of subcarriers.

\subsection{Proposed Dual Purpose SSB}
Although the \ac{UE} is capable of processing the wider bandwidth spanning multiple \ac{PRB}s in the conventional \ac{SSB}, \ac{IoT} devices cannot do so owing to their narrower receiver bandwidth and limited processing capability. Therefore, the proposed design aims to enable low-capability devices to achieve synchronization using only a portion of the \ac{SSB}, while preserving conventional synchronization for the \ac{UE} over the full \ac{SSB}. The proposed \ac{DPSSB} preserves the standard \ac{SSB} structure and the original \ac{PSS}/\ac{SSS} allocations, while selectively restructuring a subset of the \ac{DMRS} symbols. Initially, $K$ \ac{DMRS} symbols in the second and fourth \ac{OFDM} symbols denoted by $\hat{\ell}=\{1,3\}$ are relocated to a contiguous \ac{FD} region forming a single or multiple \ac{PRB}s, as illustrated in \figurename~\ref{fig_2}(b), the set of these \ac{DMRS} symbols indices is
\begin{equation}
\widetilde{\mathcal{D}}_{K}=\left\{\tilde k_{\mathrm {D}}^{\hat{\ell}}(0),
\tilde k_{\mathrm{D}}^{\hat{\ell}}(1),\ldots,\tilde k_{\mathrm {D}}^{\hat{\ell}}(K-1)\right\},
\label{eq:new_dmrs_set}
\end{equation}
where $\tilde k_{\mathrm {D}}(.)$ denotes the subcarrier relocation, the corresponding vector form notation
\begin{equation}
\mathbf{d}_{\hat{\ell},K}=
\left[d_{\mathrm{D}}^{\hat{\ell}}[m_0],
d_{\mathrm{D}}^{\hat{\ell}}[m_1],
\ldots,d_{\mathrm{D}}^{\hat{\ell}}[m_{K-1}]
\right]^{\mathrm T}\in\mathbb{C}^{K\times1}.
\label{eq:original_dmrs_vector}
\end{equation}
The \ac{PBCH} symbols displaced by this operation are remapped within the selected region through $\tilde k_{\mathrm {P}}(.)$ without changing the total number of occupied \ac{RE}s as 
\begin{equation}
\widetilde{\mathcal{P}}_{K}=\left\{\tilde k_{\mathrm {P}}^{\hat{\ell}}(0),
\tilde k_{\mathrm{P}}^{\hat{\ell}}(1),\ldots,\tilde k_{\mathrm {P}}^{\hat{\ell}}(Q_K-1)\right\},
\label{eq:new_pbch_set}
\end{equation}
where $Q_K$ denotes the total number of the remapped \ac{PBCH} symbols, and in vector form they can be expressed as
\begin{equation}
\mathbf{p}_{\hat{\ell},K}
=\left[d_{\mathrm{P}}^{\hat{\ell}}[q_0],
d_{\mathrm{P}}^{\hat{\ell}}[q_1],
\ldots,d_{\mathrm{P}}^{\hat{\ell}}[q_{Q_K-1}]
\right]^{\mathrm T}\in\mathbb{C}^{Q_K\times1}.
\label{eq:original_pbch_vector}
\end{equation}
After re-allocation, the selected \ac{DMRS} symbols, originally distributed across multiple \ac{PRB}s, are remapped into one or more contiguous \ac{PRB}(s) containing \ac{DMRS} only, while their original vacated positions are replaced by the displaced \ac{PBCH} symbols. The first subcarrier of this modified region is denoted by $k_{\mathcal{D}}^{\mathrm{Start}}
=\min \widetilde{\mathcal{D}}_{K}$, and the relocated \ac{DMRS} mapping is defined as $\tilde{k}_{\mathrm{D}}(m_k)=
k_{\mathcal{D}}^{\mathrm{Start}}+m_k$, for   $k= 0,\ldots,K-1$. Since resource allocation is performed on a per-\ac{PRB} basis (12 REs per \ac{PRB}), $k_{\mathcal{D}}^{\mathrm{Start}}$ is constrained to \ac{PRB} boundaries, i.e., $k_{\mathcal{D}}^{\mathrm{Start}}\in\{12,24,36,\ldots\}$. Accordingly, the displaced \ac{PBCH} symbols are placed immediately after and before the contiguous \ac{DMRS} block. The \ac{PBCH} modified mapping is defined as
\begin{equation}
\tilde k_{\mathrm {P}}(q_k)=
\begin{cases}
q_{k}, & k=0, \cdots,k_{\mathcal{D}}^{\mathrm{Start}}-1,
\\q_{k}+k_{\mathcal{D}}^{\mathrm{Start}}+K,&
k=0,\ldots,Q_K-1.
\end{cases}
\label{eq:pbch_destination_mapping}
\end{equation}
The remapped $\hat{\ell}$-th \ac{OFDM} symbol is therefore written as
\begin{equation}
\tilde{\text{x}}_{\hat{\ell},K}^{R}[k]
=
\begin{cases}
d_{\mathrm {D}}^{(\hat{\ell})}[m_k],&
k=\tilde{k}_{\mathrm {D}}(m_k),
\\d^{(\hat{\ell})}_{\mathrm {P}}[q_k],&
k=\tilde k_{\mathrm {P}}(q_k),\\\text{x}_{\hat{\ell}}[k],
&k\in\mathcal{A}_{K},
\end{cases}
\label{eq:remapped_symbol}
\end{equation}
where $\widetilde{\mathbf{x}}_{\hat{\ell},K}^{R}=\left[\tilde{\text{x}}_{\hat{\ell},K}^{R}[0],\ldots,\tilde{\text{x}}_{\hat{\ell},K}^{R}[N_c-1]\right]^{\mathrm {T}}$ denotes the $\ell$-th symbol after \ac{DMRS}/\ac{PBCH} remapping and $\mathcal{A}_{K}$ is expressed as
\begin{equation}
\mathcal{A}_{K}=\left\{k\in\{0,1,...,N_c-1\}: k\notin\widetilde{\mathcal{D}}_{K}\cup\widetilde{\mathcal{P}}_{K}\right\},
\label{eq:unmodified_res}
\end{equation}
which represents the unmodified \ac{RE}s placement. Furthermore, the \ac{DPSSB} resource grid after the \ac{FD} relocation, is expressed as $\mathbf{X}_{K}^{R}
=\left[\mathbf{x}_0,\,\widetilde{\mathbf {x}}_{1,K}^{R},\,\mathbf {x}_2,\,\widetilde{\mathbf {x}}_{3,K}^{R}\right]\in\mathbb{C}^{N_c\times4}$, as illustrated in \figurename~\ref{fig_2}(b). The relocated \ac{DMRS} coefficients are subsequently optimized to obtain a \ac{LPSS} in the following subsection.
\begin{figure}[!t]
\centering
\includegraphics[width=3.5in]{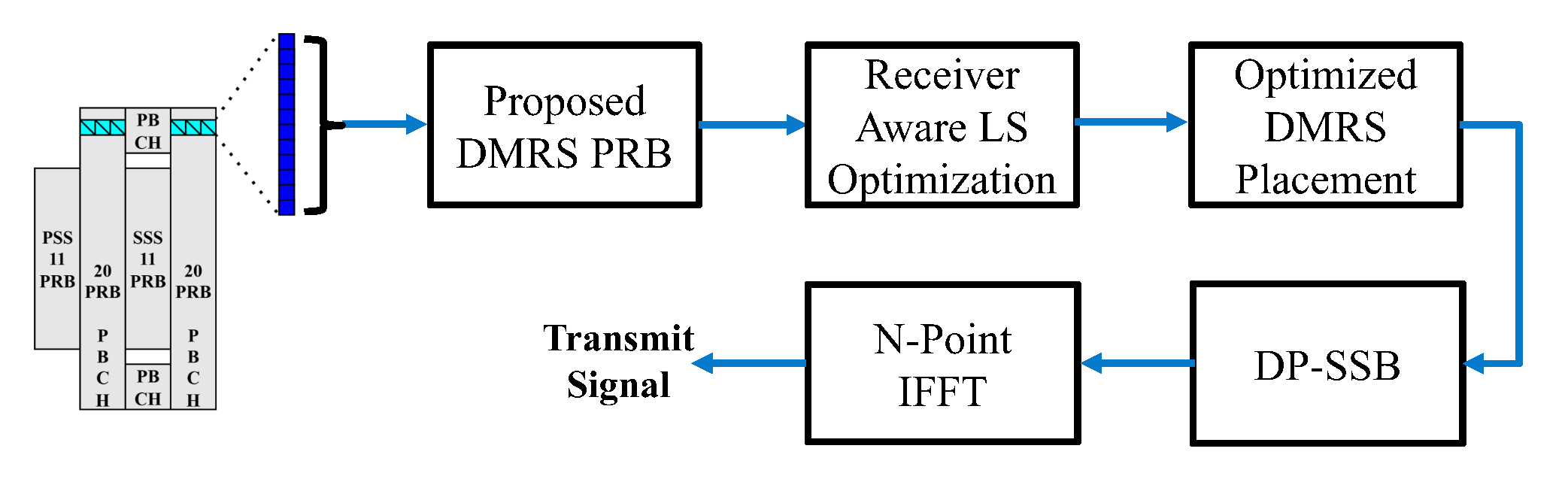}
\caption{Block diagram of the proposed \ac{LS}-based \ac{DMRS} waveform shaping and \ac{DPSSB} generation.}
\label{fig_3}
\end{figure}
\subsection{LS-Based DMRS Signal Shaping}
The contiguous \ac{DMRS} allocation introduced previously provides a set of $K$ \ac{FD} coefficients through which the \ac{LPSS} can be synthesized in the \ac{TD}. A related \ac{LS} waveform-approximation formulation using a truncated \ac{IDFT} matrix over a selected set of \ac{OFDM} subcarriers is performed, as described in \cite{11048969}. To these \ac{DMRS} subcarriers, the same underlying \ac{LS} principle is applied, which are the only \ac{RE}s the design is free to modify, whereas the \ac{PBCH} and the remaining \ac{SSB} \ac{RE}s are fixed and enter the problem as a known perturbation of the observed waveform rather than as additional degrees of freedom.\\
In particular, a selection matrix for the relocated-\ac{DMRS} is designed as
\begin{equation}
\mathbf{S}_K
=\left[\mathbf{e}_{\tilde{k}_{\mathrm{D}}(0)},
\mathbf {e}_{\tilde {k}_{\mathrm {D}}(1)},
\ldots,\mathbf{e}_{\tilde {k}_{\mathrm {D}}(K-1)}
\right]\in\{0,1\}^{N_c\times K},
\label{eq:selection_matrix}
\end{equation}
where $\mathbf {e}_{\tilde {k}_{\mathrm {D}}(k)}\in\mathbb{R}^{N_c}$ is the $k$-th canonical basis vector. Hence, the relocated DMRS symbols are extracted from the remapped OFDM symbols as $\mathbf{d}_{\hat{\ell},K}=\mathbf{S}_K^{\mathrm {T}} \widetilde{\mathbf {x}}_{\hat{\ell},K}^{R}\in\mathbb {C}^{K\times1}$, which serves as the initial value of the design variable $\mathbf{d}\in\mathbb{C}^{K\times1}$.The composition $\bar{\mathbf{E}}\triangleq\mathbf{E}\mathbf{S}_{K}\in\{0,1\}^{N\times K}$ acts as the zero-padding operator that places the $K$ relocated DMRS coefficients onto their physical $N$-point IDFT bins, so that $\bar{\mathbf{d}}_{\mathrm{D}}^{(\hat{\ell})}=\bar{\mathbf{E}}\mathbf{d}$ and, since $\mathbf{E}^{\mathrm{T}}\mathbf{E}=\mathbf{I}_{N_{c}}$ and $\mathbf{S}_{K}^{\mathrm{T}}\mathbf{S}_{K}=\mathbf{I}_{K}$, it has orthonormal columns, i.e, $\bar{\mathbf{E}}^{\mathrm{H}}\bar{\mathbf{E}}=\mathbf{I}_{K}$. Accordingly, the truncated IDFT matrix restricted to the relocated-DMRS subcarriers (given that $\bar{\mathbf{E}}$ is a column-selection matrix) is defined as
\begin{equation}
\mathbf{F}_{\mathrm{D},K}=\mathbf{F}_{N}\bar{\mathbf{E}}\in\mathbb{C}^{N\times K},\quad \mathbf{F}_{\mathrm{D},K}^{\mathrm{H}}\mathbf{F}_{\mathrm{D},K}=\mathbf{I}_{K},
\label{eq:FD_orthogonality}
\end{equation}
where its columns are the $K$ mutually orthonormal IDFT columns associated with sorting $\widetilde{\mathcal{D}}_{K}$. This formulation is comparable to the truncated representation in \cite{11048969}, with the distinction that the designable subspace herein consists exclusively of the relocated DMRS subcarriers. Therefore, the isolated \ac{TD} contribution of DMRS coefficients is 
\begin{equation}
\mathbf{s}_{\hat{\ell},K}^{\mathrm{D}}=
\mathbf{F}_{\mathrm{D},K}\mathbf{d}
\in\mathbb{C}^{N\times1}.
\label{equ:original_dmrs_td}
\end{equation}
Since the same $N$-point IDFT is applied jointly to all \ac{RE}s of the $\hat{\ell}$-th symbol, $\mathbf{s}_{\hat{\ell},K}^{\mathrm{D}}$ is never synthesized in isolation. Defining the selection matrices of the remapped PBCH and of the unmodified \ac{RE}s as
\begin{subequations}
    \begin{align}
    \mathbf{S}_K^{\mathrm{P}} &=\left[\mathbf{e}_{\tilde{k}_{\mathrm{P}}(0)},\ldots,\mathbf{e}_{\tilde{k}_{\mathrm{P}}(Q_K-1)}\right]\in\{0,1\}^{N_c\times Q_K}, \\
    \mathbf{S}_K^{\mathcal{A}}& = \left[\mathbf{e}_{k}\right]_{k\in\mathcal{A}_K}\in\{0,1\}^{N_c\times|\mathcal{A}_K|},
    \end{align}
    \label{eq:selection_matrix_pbch}
\end{subequations}
with $\mathbf{p}_{\hat{\ell},K}=\mathbf{S}_K^{\mathrm{P,T}}\widetilde{\mathbf{x}}_{\hat{\ell},K}^{R}$ and $\mathbf{x}_{\hat{\ell}}^{\mathcal{A}}=\mathbf{S}_K^{\mathcal{A},\mathrm{T}}\widetilde{\mathbf{x}}_{\hat{\ell},K}^{R}$, related with the associated truncated IDFT matrices $\mathbf{F}_{\mathrm{P},K}\triangleq\mathbf{F}_{N}\mathbf{E}\mathbf{S}_{K}^{\mathrm{P}}$ and $\mathbf{F}_{\mathcal{A},K}\triangleq\mathbf{F}_{N}\mathbf{E}\mathbf{S}_{K}^{\mathcal{A}}$. As she sets $\widetilde{\mathcal{D}}_{K}$, $\widetilde{\mathcal{P}}_{K}$ and $\mathcal{A}_{K}$ partition $\mathcal{K}_{\mathrm{SSB}}$, the transmitted time-domain symbol is
\begin{equation}
\mathbf{s}_{\hat{\ell},K}=\mathbf{F}_{\mathrm{D},K}\mathbf{d}+\underbrace{\mathbf{F}_{\mathrm{P},K}\mathbf{p}_{\hat{\ell},K}+\mathbf{F}_{\mathcal{A},K}\mathbf{x}_{\hat{\ell}}^{\mathcal{A}}}_{\triangleq\;\mathbf{z}_{\hat{\ell},K}},
\label{eq:full_symbol_td}
\end{equation}
where $\mathbf{z}_{\hat{\ell},K}\in\mathbb{C}^{N\times1}$ collects the contributions that the design absorbs.\\
$\mathbf{Remark \text{ 1}:}$ the \ac{IoT} device does not observe \eqref{eq:full_symbol_td} directly. Consistently with the low-power architectures in \cite{sahin2018sequence,zegrar2026ofdm}, it applies a narrowband filter matched to the relocated-DMRS band $\widetilde{\mathcal{D}}_{K}$, and then decimates the result to the $M$ samples processed by its correlator. Denoting  $\mathbf{H}_{f}\in\mathbb{C}^{N\times N}$ as the Toeplitz matrix of the linear convolution with the length-$L_{f}$ filter $\mathbf{h}_{f}$, and the decimation matrix $\mathbf{T}_{M}\in\{0,1\}^{M\times N}$, the front-end operator is defined by
\begin{equation}
\boldsymbol{\Phi}_{K}=\mathbf{T}_{M}\mathbf{H}_{f}\in\mathbb{C}^{M\times N},
\label{eq:frontend}
\end{equation}
and the corresponding observed waveform is
\begin{equation}
\mathbf{y}_{\hat{\ell},K}=\boldsymbol{\Phi}_{K}\mathbf{s}_{\hat{\ell},K}=\mathbf{G}_{K}\mathbf{d}+\mathbf{i}_{\hat{\ell},K},
\label{eq:observed_waveform}
\end{equation}
with the filtered signal and interference defined as
\begin{equation}
\mathbf{G}_{K}\triangleq\boldsymbol{\Phi}_{K}\mathbf{F}_{\mathrm{D},K}\in\mathbb{C}^{M\times K},\qquad \mathbf{i}_{\hat{\ell},K}\triangleq\boldsymbol{\Phi}_{K}\mathbf{z}_{\hat{\ell},K}\in\mathbb{C}^{M\times1}.
\label{eq:GK_iK}
\end{equation}
Note that $\boldsymbol{\Phi}_{K}$ is not diagonalized by the DFT matrix, where $\mathbf{H}_{f}$ has a linear convolution and $\mathbf{T}_{M}$ truncates in time. Consequently,  $\mathbf{R}\triangleq\mathbf{F}_{N}^{\mathrm{H}}\boldsymbol{\Phi}_{K}^{\mathrm{H}}\boldsymbol{\Phi}_{K}\mathbf{F}_{N}$ is not diagonal, and its off-diagonal block indexed by $\widetilde{\mathcal{D}}_{K}\times(\widetilde{\mathcal{P}}_{K}\cup\mathcal{A}_{K})$ is  non-zero, such as
\begin{equation}
\mathbf{G}_{K}^{\mathrm{H}}\mathbf{i}_{\hat{\ell},K}=(\mathbf{E}\mathbf{S}_{K})^{\mathrm{H}}\,\mathbf{R}\,\mathbf{E}\left(\mathbf{S}_{K}^{\mathrm{P}}\mathbf{p}_{\hat{\ell},K}+\mathbf{S}_{K}^{\mathcal{A}}\mathbf{x}_{\hat{\ell}}^{\mathcal{A}}\right)\neq\mathbf{0}.
\label{eq:nonorthogonality}
\end{equation}
That means that the finite stopband attenuation, the linear-convolution edge transient, and the time decimation all leak \ac{PBCH} and adjacent-\ac{PRB} energy into the band and the observation window in which the \ac{LPSS} is synthesized. This leakage is deterministic and known at the transmitter, and it must therefore be absorbed by the design.
\begin{figure*}[!t]
\centering   \includegraphics[width=0.99\linewidth]{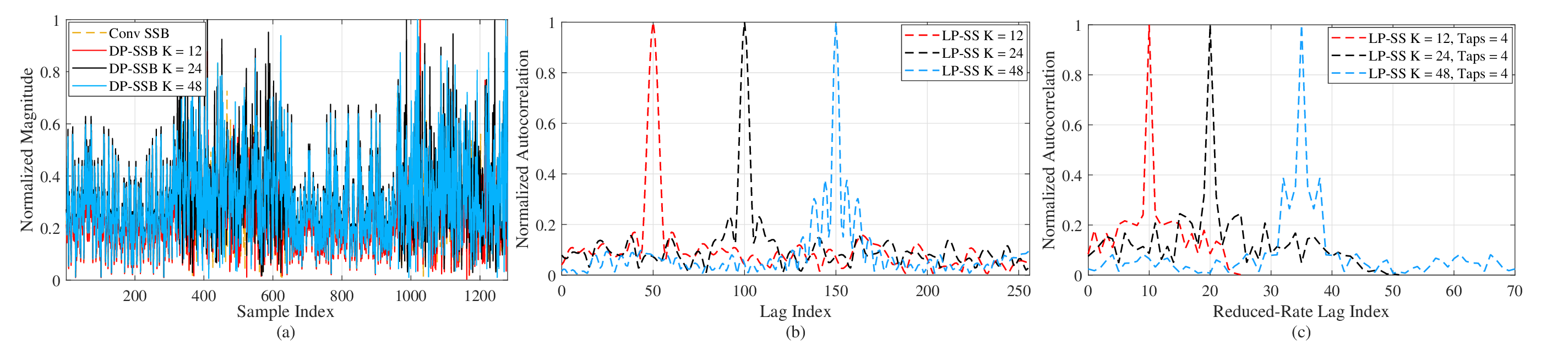}
    \caption{\ac{LPSS} design comparison for different DMRS-block sizes. (a) Transmitted \ac{DPSSB}. (b) Proposed \ac{LPSS} (filtered DP-SSB). (c) Proposed \ac{LPSS} (filtered DP-SSB) under 4-tap channel effect.}
 \label{fig_SSdeisng}
\end{figure*}
Unlike \ac{OOK}-based wake-up signals, where the targeted signal is a real on--off envelope $b^{\mathrm{tar}}[n]\in\{0,1\}$ recovered non-coherently by an envelope detector so that the subcarrier phases are unconstrained \cite{sahin2018sequence}, the desired \ac{LPSS} is intended for correlation-based \ac{TF} synchronization. Therefore, the targeted signal is a complex-valued, time-localized waveform whose phase must be preserved, since the \ac{IoT} device correlates the observation against a stored replica. To construct the desired waveform at the appropriate OFDM sampling rate, the $K$ relocated DMRS symbols are expanded to $N$ samples. Define $q_K[n]=\left\lfloor nK/N\right\rfloor\in\{0,\ldots,K-1\}$ as the DMRS-to-sample index mapping, and let $\boldsymbol{\Psi}_{K}\in\{0,1\}^{N\times K}$ be the corresponding hold matrix, $[\boldsymbol{\Psi}_{K}]_{n,k}=1$ when $k=q_{K}[n]$, so that the $N$-point reference waveform is
\begin{equation}    
\mathbf{b}_{\hat{\ell},K}^{\mathrm{ref}}=\boldsymbol{\Psi}_{K}\mathbf{d}_{\hat{\ell},K},\qquad b_{\hat{\ell},K}^{\mathrm{ref}}[n]=d_{\mathrm{D}}^{(\hat{\ell})}\!\left[q_{K}[n]\right]. \label{eq:reference_waveform}
\end{equation}
Each relocated DMRS symbol thus occupies $N/K$ consecutive samples, which plays the role of the chip duration $L=N/M$ of the \ac{OOK} construction in \cite{zegrar2026ofdm}, the essential difference being that the chips here are a combination of complex DMRS symbols rather than binary amplitudes. To concentrate the magnitude of the reference waveform in the \ac{TD}, it is weighted by a Gaussian window $\bar{g}_K[n]=e^{-(n-\mu_K)^2/2\sigma_K^2}$ normalized to unit peak as
\begin{equation}
        g_K[n]=  \frac{\bar{g}_K[n]}{\max_{n} \bar{g}_K[n]},
\end{equation}
where $\mu_K=\frac{N}{\rho_K}$, $\rho_K$ controls the desired peak concentration, and $\sigma_K$ determines the temporal spread. In vector form $\mathbf g_K = [g_K[0],\ldots,g_K[N-1]]^{\mathrm T}$, and the target waveform is
\begin{equation}
\mathbf{b}_{\hat{\ell},K}^{\mathrm{tar}}=
\mathbf{b}_{\hat{\ell},K}^{\mathrm{ref}}\odot\mathbf{g}_K
=\mathrm{diag}(\mathbf{g}_{K})\boldsymbol{\Psi}_{K}\mathbf{d}_{\hat{\ell},K}
\in\mathbb{C}^{N\times1}.
\label{eq:target_waveform1}
\end{equation}
Since the \ac{DMRS} symbols are drawn from a constant-modulus alphabet, $|d_{\mathrm{D}}^{(\hat{\ell})}[m]|=\varepsilon_{\mathrm{D}}$, $\mathbf{b}_{\hat{\ell},K}^{\mathrm{tar}}$ decomposes into
\begin{equation}
\left|b_{\hat{\ell},K}^{\mathrm{tar}}[n]\right|=\varepsilon_{\mathrm{D}}\,g_{K}[n],\qquad \arg b_{\hat{\ell},K}^{\mathrm{tar}}[n]=\arg d_{\mathrm{D}}^{(\hat{\ell})}\!\left[q_{K}[n]\right],
\label{eq:mag_phase_split}
\end{equation}
which means that the magnitude is a pure Gaussian pulse of width $\sigma_{K}$ centered at $\mu_{K}$, which provides the sharp time-domain peak, while the phase carries the pseudo-random Gold-sequence content of the \ac{DMRS} and hence the cell-specific and \ac{SSB}-index information. This is precisely the property that an envelope-detection \ac{OOK} receiver would discard \cite{zegrar2026ofdm}, and the reason why the present design must be formulated over both magnitude and phase constraints.
\begin{algorithm}[!t]
    \caption{LS-Based DP-SSB Signal Shaping}
    \SetAlgoLined
    \SetKwInOut{Input}{Input}
    \SetKwInOut{Output}{Output}
    \DontPrintSemicolon
    \Input{$\widetilde{\mathbf{x}}_{1,K}^{R}$, $\widetilde{\mathbf{x}}_{3,K}^{R}$, $K$,  $\boldsymbol{\Phi}_{K}$, $\rho_{K}$, $\sigma_{K}$}
    \Output{$\widetilde{\mathbf{x}}_{1,K}^{\mathrm{DP}}$ and $\widetilde{\mathbf{x}}_{3,K}^{\mathrm{DP}}$}
    Construct $\mathbf{F}_{\mathrm{D},K}\rightarrow\mathbf{F}_{N}\mathbf{E}\mathbf{S}_{K}$, $\mathbf{G}_{K}\rightarrow\boldsymbol{\Phi}_{K}\mathbf{F}_{\mathrm{D},K}$, $\boldsymbol{\Psi}_{K}$, $\mathbf{g}_{K}$\\
    \For{$\hat{\ell} \in \{1,3\}$}
    {
    Extract $\mathbf{d}_{\hat{\ell},K}$ from $\widetilde{\mathbf{x}}_{\hat{\ell},K}^{R}$\\
    Construct $\mathbf{b}_{\hat{\ell},K}^{\mathrm{ref}}\leftarrow\boldsymbol{\Psi}_{K}\mathbf{d}_{\hat{\ell},K}$ and $\mathbf{b}_{\hat{\ell},K}^{\mathrm{tar}}\leftarrow\mathbf{b}_{\hat{\ell},K}^{\mathrm{ref}}\odot\mathbf{g}_{K}$\\
    Evaluate the filtered interference $\mathbf{i}_{\hat{\ell},K}$\\
    Form the effective target $\widetilde{\mathbf{b}}_{\hat{\ell},K}^{\mathrm{tar}}\rightarrow\mathbf{b}_{\hat{\ell},K}^{\mathrm{tar}}-\mathbf{i}_{\hat{\ell},K}$\\
    Solve the LS problem $\bar{\mathbf{d}}_{\hat{\ell},K}\rightarrow\mathbf{G}_{K}^{\dagger}\widetilde{\mathbf{b}}_{\hat{\ell},K}^{\mathrm{tar}}$\\
    Normalize $\bar{\mathbf{d}}_{\hat{\ell},K}$ to obtain $\mathbf{d}_{\hat{\ell},K}^{\star}$\\
    Reinsert $\mathbf{d}_{\hat{\ell},K}^{\star}$ into $\widetilde{\mathbf{x}}_{\hat{\ell},K}^{R}$ to obtain $\widetilde{\mathbf{x}}_{\hat{\ell},K}^{\mathrm{DP}}$
    }
 \label{LS_final}
\end{algorithm}
\par Consequently, the synchronization of the designed waveform is assessed by its periodic autocorrelation. For $\mathbf{s}_{\hat{\ell},K}^{\mathrm{D}}=\mathbf{F}_{\mathrm{D},K}\mathbf{d}$, the circular autocorrelation property of the $N$-point DFT grid yields
\begin{equation}
r_{\hat{\ell},K}[\tau]=\!\!\sum_{n=0}^{N-1}\!\! s_{\hat{\ell},K}^{\mathrm{D}}[n]\,s_{\hat{\ell},K}^{\mathrm{D},H}\big[(n-\tau)_{N}\big] =\!\!\!\sum_{k\in\widetilde{\mathcal{D}}_{K}}\!\!\!\left|\bar{d}_{\mathrm{D}}^{(\hat{\ell})}[k]\right|^{2}\!e^{j2\pi k\tau/N},
\label{eq:autocorrelation11}
\end{equation}
which implies that the autocorrelation depends only on the magnitude profile of the DMRS coefficients, whereas their phase profile fixes the position of the peak in time. Therefore, replacing $\bar{d}_{\mathrm{D}}^{(\hat{\ell})}[k]$ by $\bar{d}_{\mathrm{D}}^{(\hat{\ell})}[k]e^{-j2\pi kn_{0}/N}$ shifts $\mathbf{s}_{\hat{\ell},K}^{\mathrm{D}}$ by $n_{0}$ samples while keeping $r_{\hat{\ell},K}[\tau]$ unchanged. In the constant-modulus limit $|\bar{d}_{\mathrm{D}}^{(\hat{\ell})}[k]|^{2}=E_{\hat{\ell},K}^{\mathrm{D}}/K$ over the contiguous block $\widetilde{\mathcal{D}}_{K}$, with $E_{\hat{\ell},K}^{\mathrm{D}}$ denoting the total energy of the relocated DMRS block, \eqref{eq:autocorrelation11} reduces to 
\begin{equation}
\left|r_{\hat{\ell},K}[\tau]\right|=\frac{E_{\hat{\ell},K}^{\mathrm{D}}}{K}\left|\frac{\sin\!\left(\pi K\tau/N\right)}{\sin\!\left(\pi \tau/N\right)}\right|,
\label{eq:dirichlet}
\end{equation}
whose mainlobe half-width is $N/K$ samples. Hence, the temporal resolution of the \ac{LPSS} improves linearly with $K$, which explains the sharpening observed in \figurename~\ref{fig_SSdeisng}(b) as $K$ grows. Therefore, Gaussian weighting reflects an explicit trade-off: in the \ac{FD} it convolves the reference signal with a kernel of width $\propto N/\sigma_{K}$, such that a small $\sigma_{K}$ sharpens the \ac{TD} peak but tapers $|\bar{d}_{\mathrm{D}}^{(\hat{\ell})}[k]|$ across $\widetilde{\mathcal{D}}_{K}$, raising the correlation sidelobes in \eqref{eq:autocorrelation11} and increasing the energy that falls outside the $K$ designable bins, whereas a large $\sigma_{K}$ restores a flat, constant-envelop like profile at the cost of a broader peak \cite{chu1972polyphase}.
\par Subsequently, the design must reproduce $\mathbf{b}_{\hat{\ell},K}^{\mathrm{tar}}$ where the fixed contribution $\mathbf{i}_{\hat{\ell},K}$ is unavoidably superimposed. Consequently, the \ac{LS} problem becomes
\begin{equation}
\bar{\mathbf{d}}_{\hat{\ell},K}
=\arg\min_{\mathbf{d}}
\underbrace{\left\|\mathbf{G}_{K}\mathbf{d}
-\widetilde{\mathbf{b}}_{\hat{\ell},K}^{\mathrm{tar}}
\right\|_2^2}_{\mathcal{J}(\mathbf{d})},
\qquad
\widetilde{\mathbf{b}}_{\hat{\ell},K}^{\mathrm{tar}}=\mathbf{b}_{\hat{\ell},K}^{\mathrm{tar}}-\mathbf{i}_{\hat{\ell},K},
\label{eq:ls_problem}
\end{equation}
where $\widetilde{\mathbf{b}}_{\hat{\ell},K}^{\mathrm{tar}}$ is the desired shape after absorbing the deterministic impact of the \ac{PBCH} and unmodified \ac{RE}s. Expanding the cost function gives
\begin{equation}
\begin{aligned}
\mathcal{J}(\mathbf{d})
=\mathbf{d}^{\mathrm{H}}\mathbf{G}_{K}^{\mathrm{H}}\mathbf{G}_{K}\mathbf{d}
-2\Re\left\{\mathbf{d}^{\mathrm{H}}
\mathbf{G}_{K}^{\mathrm {H}}
\widetilde{\mathbf{b}}_{\hat{\ell},K}^{\mathrm{tar}}
\right\}+\left\|
\widetilde{\mathbf{b}}_{\hat{\ell},K}^{\mathrm{tar}}
\right\|_2^2,
\end{aligned}
\label{eq:ls_expansion}
\end{equation}
and setting the gradient with respect to $\mathbf{d}^{\mathrm{H}}$ to zero,
\begin{equation}
        \frac{\partial \mathcal{J}(\mathbf{d})}{\partial \mathbf{d}^{\mathrm{H}}}= \mathbf{G}_{K}^\mathrm{H}\mathbf{G}_{K} \mathbf{d}-\mathbf{G}_{K}^\mathrm{H} \widetilde{\mathbf{b}}_{\hat{\ell},K}^{\mathrm{tar}}=\mathbf{0},
\label{eq:ls_gradient}
\end{equation}
for $\mathbf{G}_{K}$ of full column rank ($M\ge K$), the LS solution is
\begin{equation}
\bar{\mathbf{d}}_{\hat{\ell},K}=\mathbf{G}_{K}^{\dagger}\widetilde{\mathbf{b}}_{\hat{\ell},K}^{\mathrm{tar}}
=\underbrace{\mathbf{G}_{K}^{\dagger}\mathbf{b}_{\hat{\ell},K}^{\mathrm{tar}}}_{\text{waveform shaping}}-\underbrace{\mathbf{G}_{K}^{\dagger}\mathbf{i}_{\hat{\ell},K}}_{\text{interference}},
\label{eq:ls_simplified}
\end{equation}
with $\mathbf{G}_{K}^{\dagger}=(\mathbf{G}_{K}^{\mathrm{H}}\mathbf{G}_{K})^{-1}\mathbf{G}_{K}^{\mathrm{H}}$. The interference term in \eqref{eq:ls_simplified} is the contribution that the earlier \ac{DMRS}-only formulation ignores, where it uses the $K$ relocated \ac{DMRS} coefficients to pre-compensate the portion of the \ac{PBCH} and residual-\ac{SSB} leakage that falls inside the designable subspace.
\par Let $\mathbf{P}_{K}=\mathbf{G}_{K}\mathbf{G}_{K}^{\dagger}$ denote the orthogonal projector onto the achievable subspace 
\begin{equation}
    \begin{aligned}
    \operatorname{range}\left(\mathbf{G}_K\right)=\left\{\mathbf{G}_K \mathbf{d}: \mathbf{d} \in \mathbb{C}^K\right\} \subset \mathbb{C}^M,
    \end{aligned}
\end{equation}
and let
\begin{equation}
\mathbf{r}_{\hat{\ell},K}(\mathbf{d})=\mathbf{y}_{\hat{\ell},K}-\mathbf{b}_{\hat{\ell},K}^{\mathrm{tar}}=\mathbf{G}_{K}\mathbf{d}+\mathbf{i}_{\hat{\ell},K}-\mathbf{b}_{\hat{\ell},K}^{\mathrm{tar}}\in\mathbb{C}^{M\times1}
\label{eq:residual_def}
\end{equation}
denotes the residual observed at the correlator. Evaluating \eqref{eq:residual_def} for the interference-aware solution $\bar{\mathbf{d}}_{\hat{\ell},K}$ of \eqref{eq:ls_simplified} and for the uncompensated design $\mathbf{G}_{K}^{\dagger}\mathbf{b}_{\hat{\ell},K}^{\mathrm{tar}}$ gives
\begin{subequations}
\begin{align}
\mathbf{r}^{\mathrm{comp}}_{\hat{\ell},K}&=-\left(\mathbf{I}_{M}-\mathbf{P}_{K}\right)\widetilde{\mathbf{b}}_{\hat{\ell},K}^{\mathrm{tar}},\\
\mathbf{r}^{\mathrm{unc}}_{\hat{\ell},K}&=-\left(\mathbf{I}_{M}-\mathbf{P}_{K}\right)\mathbf{b}_{\hat{\ell},K}^{\mathrm{tar}}+\mathbf{i}_{\hat{\ell},K}
=\mathbf{r}^{\mathrm{comp}}_{\hat{\ell},K}+\mathbf{P}_{K}\mathbf{i}_{\hat{\ell},K},
\end{align}
\label{eq:two_residuals}
\end{subequations}
where $\mathbf{r}^{\mathrm{comp}}_{\hat{\ell},K}$ and $\mathbf{r}^{\mathrm{unc}}_{\hat{\ell},K}$ denote the resulting residuals for the compensated design of \eqref{eq:ls_simplified} and for the uncompensated design $\mathbf{G}_{K}^{\dagger}\mathbf{b}_{\hat{\ell},K}^{\mathrm{tar}}$, respectively. Since $\mathbf{r}^{\mathrm{comp}}_{\hat{\ell},K}\in\mathrm{range}(\mathbf{G}_{K})^{\perp}$ while $\mathbf{P}_{K}\mathbf{i}_{\hat{\ell},K}\in\mathrm{range}(\mathbf{G}_{K})$, the two terms in \eqref{eq:two_residuals} are orthogonal and
\begin{equation}
\left\|\mathbf{r}^{\mathrm{unc}}_{\hat{\ell},K}\right\|_{2}^{2}
=\left\|\mathbf{r}^{\mathrm{comp}}_{\hat{\ell},K}\right\|_{2}^{2}
+\left\|\mathbf{P}_{K}\mathbf{i}_{\hat{\ell},K}\right\|_{2}^{2},
\label{eq:gain_of_compensation}
\end{equation}
which means that absorbing the interference reduces the waveform-approximation error by the energy of the leakage that is observable within the designable subspace. The remaining term $\|(\mathbf{I}_{M}-\mathbf{P}_{K})\widetilde{\mathbf{b}}_{\hat{\ell},K}^{\mathrm{tar}}\|_{2}^{2}$ is irreducible with $K$ degrees of freedom and decreases as $K$ grows. \\
$\mathbf{Remark \text{ 2}:}$ for an ideal full-length circular filter with $M=N$, $\mathbf{T}_{M}=\mathbf{I}_{N}$ and $\boldsymbol{\Phi}_{K}=\mathbf{F}_{N}\boldsymbol{\Lambda}\mathbf{F}_{N}^{\mathrm{H}}$ with $\boldsymbol{\Lambda}=\mathrm{diag}(\lambda_{0},\ldots,\lambda_{N-1})$, than $\mathbf{R}=\boldsymbol{\Lambda}^{\mathrm{H}}\boldsymbol{\Lambda}$ is diagonal. Since $\widetilde{\mathcal{D}}_{K}$, $\widetilde{\mathcal{P}}_{K}$ and $\mathcal{A}_{K}$ are disjoint, \eqref{eq:nonorthogonality} gives $\mathbf{G}_{K}^{\mathrm{H}}\mathbf{i}_{\hat{\ell},K}=\mathbf{0}$, the interference term in \eqref{eq:ls_simplified} vanishes, and \eqref{eq:ls_simplified} collapses to the decoupled solution $\bar{\mathbf{d}}_{\hat{\ell},K}=\mathbf{F}_{\mathrm{D},K}^{\mathrm{H}}\mathbf{b}_{\hat{\ell},K}^{\mathrm{tar}}$ used in \cite{11048969,zegrar2026ofdm}, which is the truncated-\ac{IDFT} form. 
Therefore, although the solution minimizes the waveform approximation error, its energy is not equal to that of the original relocated \ac{DMRS} symbols. To avoid changing the total energy assigned to the relocated \ac{DMRS} \ac{RE}s, we define $E_{\hat{\ell},K}^{\mathrm{D}}=\left\|\mathbf {d}_{\hat{\ell},K}\right\|_2^2$ and the normalization factor $\eta_{\hat{\ell},K}=\sqrt{E_{\hat{\ell},K}^{\mathrm {D}}/\|\bar{\mathbf {d}}_{\hat{\ell},K}\|_2^2}$. For $\|\bar{\mathbf {d}}_{\hat{\ell},K}\|_2>0$, the resulting optimized \ac{DMRS} vector is
\begin{equation}
\mathbf{d}_{\hat{\ell},K}^{\star}=\eta_{\hat{\ell},K}\bar{\mathbf {d}}_{\hat{\ell},K},
\label{eq:energy_normalized_dmrs}
\end{equation}
ensuring $\|\mathbf {d}_{\hat{\ell},K}^{\star}\|_2^2=\|\mathbf {d}_{\hat{\ell},K}\|_2^2$. Since \eqref{eq:energy_normalized_dmrs} scales all coefficients by the same real factor, it scales \eqref{eq:autocorrelation11} by $\eta_{\hat{\ell},K}^{2}$ and therefore leaves the normalized autocorrelation $r_{\hat{\ell},K}[\tau]/r_{\hat{\ell},K}[0]$, and its peak-to-sidelobe ratio unaltered. The \ac{LS} optimization followed by \eqref{eq:energy_normalized_dmrs} is equivalently written as
\begin{equation}
\begin{aligned}
\mathbf{d}_{\hat{\ell},K}^{\star}
=\arg\min_{\mathbf {d}}
&\left\|\mathbf{G}_{K}\mathbf {d}
-\widetilde{\mathbf {b}}_{\hat{\ell},K}^{\mathrm{tar}}\right\|_2^2
\\\mathrm{s.t.}\;&\|\mathbf {d}\|_2^2=E_{\hat{\ell},K}^{\mathrm {D}}.
\end{aligned}
\label{eq:constrained_ls}
\end{equation}
When the signal response is flat over $\widetilde{\mathcal{D}}_{K}$, so that $\mathbf{G}_{K}^{\mathrm{H}}\mathbf{G}_{K}=\alpha\mathbf{I}_{K}$ with $\alpha$ a constant, the quadratic term of \eqref{eq:ls_expansion} reduces to $\gamma E_{\hat{\ell},K}^{\mathrm{D}}$ under the energy constraint $\|\mathbf{d}\|_2^2=E_{\hat{\ell},K}^{\mathrm{D}}$, so minimizing $\mathcal{J}(\mathbf{d})$ is equivalent to maximizing $\Re\{\mathbf{d}^{\mathrm{H}}\mathbf{G}_{K}^{\mathrm{H}}\widetilde{\mathbf{b}}_{\hat{\ell},K}^{\mathrm{tar}}\}$. Consequently, this is maximized when $\mathbf{d}$ is aligned with $\mathbf{G}_{K}^{\mathrm{H}}\widetilde{\mathbf{b}}_{\hat{\ell},K}^{\mathrm{tar}}$, giving
\begin{equation}
\mathbf {d}_{\hat{\ell},K}^{\star}
=\sqrt{E_{\hat{\ell},K}^{\mathrm {D}}}\,
\frac{\mathbf{G}_{K}^{\mathrm {H}}
\widetilde{\mathbf {b}}_{\hat{\ell},K}^{\mathrm{tar}}
}{\left\|\mathbf{G}_{K}^{\mathrm {H}}
\widetilde{\mathbf {b}}_{\hat{\ell},K}^{\mathrm{tar}}\right\|_2}.
\label{eq:constrained_solution}
\end{equation}
Finally, the relocated \ac{DMRS} coefficients in the remapped \ac{OFDM} symbol are replaced by the optimized \ac{DMRS} coefficients $\mathbf{d}_{\hat{\ell},K}^{\star}$ as
\begin{equation}
\widetilde{\mathbf {x}}_{\hat{\ell},K}^{\mathrm{DP}}
=(\mathbf{I}_{N_c}-\mathbf{S}_K\mathbf{S}_K^{\mathrm{H}})\widetilde{\mathbf {x}}_{\hat{\ell},K}^{R}+\mathbf{S}_K\mathbf{d}_{\hat{\ell},K}^{\star}.
\end{equation}
The resulting \ac{DPSSB} resource grid is expressed as $\mathbf{X}_{K}^{\mathrm{DP}}
=\left[\mathbf{x}_0,\,
\widetilde{\mathbf{x}}_{1,K}^{\mathrm{DP}},\,
\mathbf{x}_2,\,\widetilde{\mathbf{x}}_{3,K}^{\mathrm{DP}}\right]\in \mathbb{C}^{N_c\times 4}$. After embedding the \ac{SSB} subcarriers onto the physical \ac{IFFT} grid via $\bar{\mathbf{X}}_K^{\mathrm{DP}}
=\mathbf{E}\mathbf{X}_K^{\mathrm{DP}}$, the time-domain \ac{DPSSB} symbols are $\mathbf{S}_K^{\mathrm{DP}}
=\mathbf{F}_N\bar{\mathbf {X}}_K^{\mathrm{DP}}
=\mathbf{F}_N\mathbf {E}\mathbf {X}_K^{\mathrm{DP}}\in \mathbb{C}^{N\times 4}$. The proposed \ac{DPSSB}, can be represented in \ac{DT} as
\begin{equation}
s_{\ell,K}^{\mathrm{DP}}[n]
=\frac{1}{\sqrt{N}}\sum_{k=0}^{N-1}\bar X_K^{\mathrm{DP}}[k,\ell]
e^{j2\pi kn/N}.
\label{eq:dp_td_cp}
\end{equation}
The overall signal shaping and design procedure is illustrated in \figurename~\ref{fig_3} and summarized in Algorithm 1.

\section{Proposed Synchronization Framework}
This section presents the proposed low-complexity detection and synchronization framework enabled via \ac{DPSSB} for low-power \ac{IoT} devices. Unlike conventional \ac{UE} receivers that can process the complete \ac{SSB} bandwidth, the proposed \ac{IoT} receiver architecture exploits the $K$-dependent \ac{DMRS} \ac{PRB} embedded in the \ac{DPSSB} to obtain a narrowband observation of the \ac{LPSS}. The proposed synchronization procedure follows a hierarchical framework including coarse and fine \ac{TF} synchronization after the \ac{DPSSB} detection, as illustrated in Fig.~\ref{fig_61}.
\begin{figure}[!t]
\centering
\includegraphics[width=3.3in]{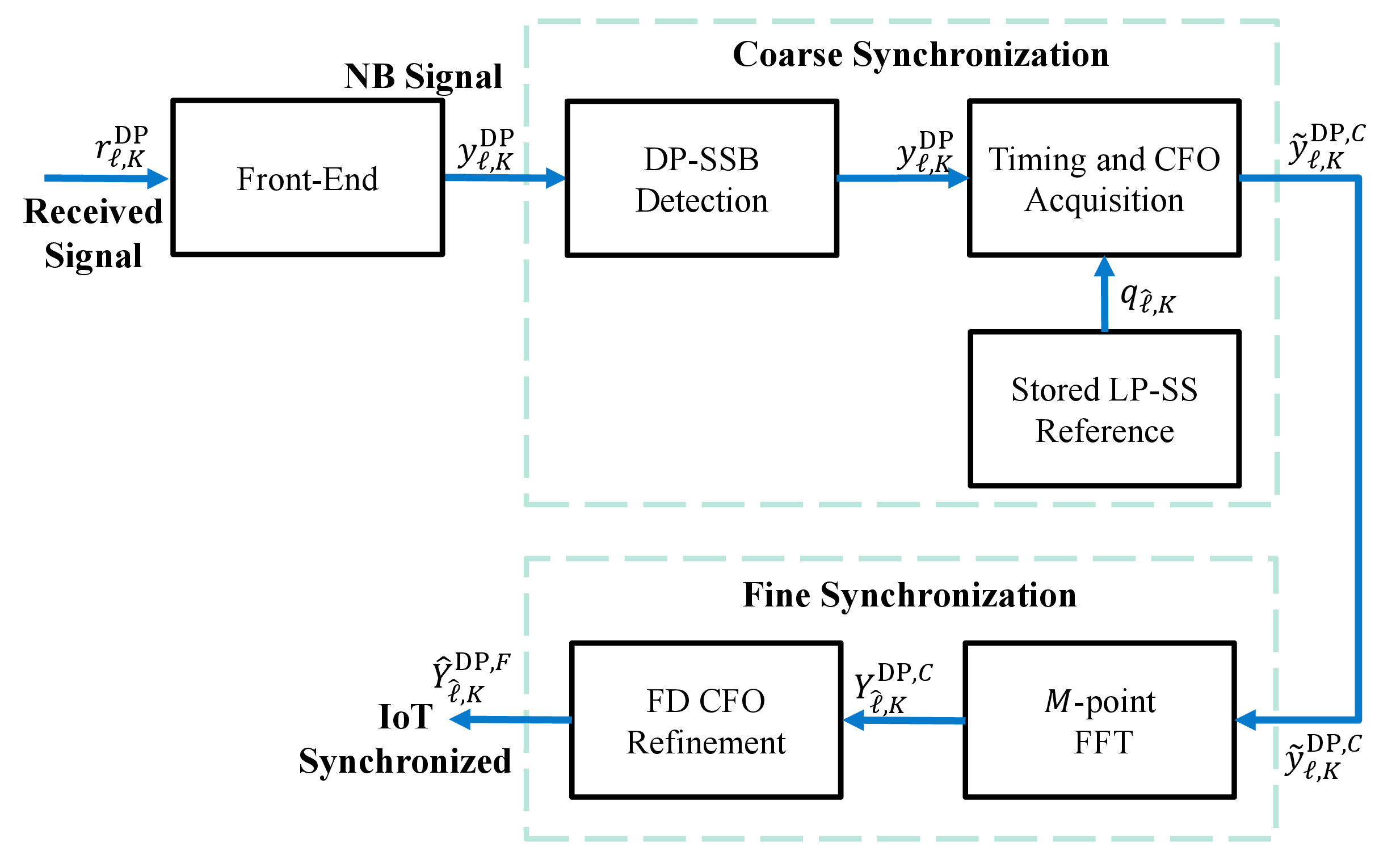}
\caption{Block diagram of the proposed \ac{TF} synchronization procedure.}
\label{fig_61}
\end{figure}
\subsection{DP-SSB Reception}
As introduced earlier in Section~II, the proposed \ac{DPSSB} is designed such that the \ac{LPSS} can be observed by the \ac{IoT} device through the low-power receiver front-end operator $\boldsymbol{\Phi}_{K}$. Unlike a conventional UE that can process the complete \ac{SSB}, the \ac{IoT} receiver must first limit the received signal to the $K$-\ac{PRB} \ac{LPSS} observable region and reduce the sampling rate through decimation. Before proceeding to these receiver operations, the effects of the wireless channel, unknown timing and frequency offsets, and the front-end transformation on the designed \ac{LPSS} must be characterized. 

Let $\tau_0\in\mathbb{Z}\geq0$ and $f_0\in\mathbb{R}$ denote the unknown integer \ac{TO} and  \ac{CFO}, respectively. The $L_c$-tap \ac{CIR} is represented by $\mathbf{h}_c=\big[h_c[0]\quad h_c[1] \dots h_c[L_c - 1]\big]^T \in \mathbb{C}^{L_c\times1}$ and $\mathbf{H}_c\in\mathbb{C}^{N\times N}$ denotes the Toeplitz matrix of the linear convolution with $\mathbf{h}_c$. Furthermore, the unknown \ac{TO} is represented by the linear delay matrix $\mathbf{P}_{\tau_0}\in\{0,1\}^{N\times N}$ with $0\leq\tau_0<N$, such that $\mathbf{P}_{\tau_0}\mathbf{s}_{\ell,K}^{\text{DP}}\in \mathbb{C}^{N\times 4}$ introduces a linear delay shift and the \ac{CFO} induced phase rotation is indicated by $\mathbf{\Omega}_{N}(f_0)=\mathrm{diag}\{e^{j2\pi f_{o}n/F_{s}}\}_{n=0}^{N-1}\in\mathbb{C}^{N\times N}$, where $F_s=N\Delta f$ is the sampling frequency and $\Delta f$ is the subcarrier frequency. Moreover, the complex \ac{AWGN} prior to filtering is modeled as $\mathbf{w}_{\ell,K}\sim\mathcal{CN}(0,\sigma_w^2\mathbf{I}_N) \in \mathbb{C}^{N\times 4}$, where $\sigma_w^2$ denotes the noise variance. The observed wideband \ac{DPSSB} signal at the input of the \ac{IoT} receiver is expressed as
\begin{equation}
\mathbf{r}_{\ell,K}^{\text{DP}}=\mathbf{\Omega}_{N}(f_0)\mathbf{H}_{c}\mathbf{P}_{\tau_0}\mathbf{s}_{\ell,K}^{\text{DP}}+\mathbf{w}_{\ell,K}\in\mathbb{C}^{N\times 4}.
\label{eq:rx_received}
\end{equation}
Furthermore, $\mathbf{r}_{\ell,K}^{\text{DP}}$ is passed through the front-end operator in \eqref{eq:frontend} for which the \ac{LPSS} is designed. Applying $\boldsymbol{\Phi}_{K}$ to \eqref{eq:rx_received}, the \ac{NB} received signal available to the low-power receiver is
\begin{equation}
    \begin{aligned}
\mathbf{y}^{\mathrm{DP}}_{\ell,K}&=\boldsymbol{\Phi}_{K}\mathbf{r}^{\mathrm{DP}}_{\ell,K},\\
&=\boldsymbol{\Phi}_{K}\mathbf{\Omega}_{N}(f_0)\mathbf{H}_{c}\mathbf{P}_{\tau_0}\mathbf{s}_{\ell,K}^{\text{DP}}+\mathbf{v}_{\ell,K}\in\mathbb{C}^{M\times 4},
\label{IoTfrontend}
    \end{aligned}    
\end{equation}
where $\mathbf{v}_{\ell,K}\triangleq\boldsymbol{\Phi}_{K}\mathbf{w}_\ell\in\mathbb{C}^{M\times 4}$ denotes the noise component after filtering and decimation. Consequently, the properties established for $\mathbf{s}^{\mathrm{DP}}_{\ell,K}$ must be re-examined in the $M$-dimensional subspace. The same conditions apply to the filtered noise, as the linear transformation preserves the complex Gaussianity; thus, $\mathbf{v}_{\ell,K}$ satisfies $\mathbf{v}_{\ell,K}\sim\mathcal{CN}(0,\mathbf{C}_{v,K})$, where $\mathbf{C}_{v,K}=\sigma_w^2\boldsymbol{\Phi}_{K}\boldsymbol{\Phi}_{K}^H\in\mathbb{C}^{M\times M}$. However, $\mathbf{C}_{v,K}$ is governed by $\boldsymbol{\Phi}_{K}\boldsymbol{\Phi}_{K}^H$ resulting in $\mathbf{v}_{\ell,K}$ not being strictly white. 
\\
The \ac{LPSS} correlation structure available to it must survive the channel, filtering, and the decimation outlined in \eqref{IoTfrontend}. Let the channel-impaired isolated \ac{LPSS} be denoted by $u_{\hat{\ell},K}[n]=\sum_{p=0}^{L_{c}-1}h_c[p]s^{\mathrm{D}}_{\hat{\ell},K}[n-p]$. The \ac{LPSS} autocorrelation shifted by the relative path delay $p-q$ is given by 
\begin{equation}
    \begin{aligned}
      r_{u,K}[\tau]&= \!\!\sum_{n=0}^{N-1}\!\! u_{\hat{\ell},K}[n]\,u_{\hat{\ell},K}^{H}\big[(n-\tau)\big],\\
&=\!\!\sum_{p=0}^{L_c-1}\sum_{q=0}^{L_c-1}\!\! h_c[p]h_c^{H}[q]r_{\hat{\ell},K}^D[\tau-(p-q)].
\label{autoco}  
    \end{aligned}
\end{equation}
In the case of $p=q$, the original correlation structure is preserved but scaled by $|h_c[p]|^2$, whereas the terms with $p\neq q$ introduce shifted path components around the main-lobe peak, as highlighted in \figurename~\ref{fig_SSdeisng}(c). Moreover, $u_{\hat{\ell},K}[n]$ is passed through the \ac{NB} filter
to obtain $z_{\hat{\ell},K}[n]=\sum_{l=0}^{L_f-1}h_f[l]u_{\hat{\ell},K}[n-l]$, which is the weighted sum of $L_f$ neighboring samples of the \ac{LPSS}. To characterize the temporal correlation structure, consider the autocorrelation of the filtered \ac{LPSS} as $r_{z,K}[\tau]=\sum_{n}z_{\hat{\ell},K}[n]z_{\hat{\ell},K}^H[n-\tau]$, substituting the terms of $z_{\hat{\ell},K}[n]$, and interchanging the summations, we get
\begin{equation}
    r_{z,K}[\tau]=\sum_{l=0}^{L_f-1}\sum_{p=0}^{L_f-1}h_f[l]h_f^{H}[p]\sum_{n}u_{\hat{\ell},K}[n-l]u_{\hat{\ell},K}^H[n-\tau-p].
    \label{filtauto}
\end{equation}
Denoting the change of variable as $q=n-l$, the \ac{LPSS} correlation can be expressed as $\sum_{n}u_{\hat{\ell},K}[n-l]u_{\hat{\ell},K}^H[n-\tau-p]=r_{u,K}[\tau-(l-p)]$, and the \eqref{filtauto} becomes 
\begin{equation}
    r_{z,K}[\tau]=\sum_{l=0}^{L_f-1}\sum_{p=0}^{L_f-1}h_f[l]h_f^{H}[p]r_{u,K}[\tau-(l-p)].
    \label{filtcorr}
\end{equation}
The \ac{NB} filter modifies the \ac{LPSS} and produces displacement according to relative filter tap separation. Furthermore, denoting the aperiodic autocorrelation of the filter by $r_{h_f}$, and substituting $\lambda=l-p$, \eqref{filtcorr} can be expressed as the convolution of autocorrelation of \ac{LPSS} with the autocorrelation of the filter 
\begin{equation}
    r_{z,K}[\tau]=r_{u,K}[\tau]*r_{h_f}[\tau].
    \label{conv}
\end{equation}
The sharp correlation structure illustrated in \figurename~\ref{fig_SSdeisng}(c), when passed through the receiver front-end, will be shaped by $r_{h_f}[\tau]$. But for a \ac{NB} filter with sufficient taps, the expected modification will be localized around the main lobe of the original correlation. This phenomenon is expressed in \figurename~\ref{fig_SSdeisng}(b), as $K$ increases, the dominant peak and the side lobes become more pronounced.

Following the \ac{NB} filtering, the sampling rate is reduced according to the bandwidth occupied by the $K$-\ac{PRB} of the \ac{LPSS}. Let $D_k=N/M$ denote the decimation factor associated with the modified sampling frequency at the output of the \ac{IoT} front-end as $F_{s,K}=F_s/D_K$. The decimation matrix $\mathbf{T}_M$ retains the $D_k$-th sample of the filtered signal and the reduced-rate \ac{LPSS} can be written as $y_{\hat{\ell},K}^{\text{DP}}[\tilde{n}]=r_{\hat{\ell},K}^{\text{DP}}[\tilde{n}D_K]$, where $\tilde{n}=0,\dots,M-1$. For analyzing the effect of decimation on the correlation structure of the \ac{LPSS}, the reduced-rate autocorrelation is expressed as $r_{y,K}[\kappa]=\sum_{\tilde{n}}y_{\hat{\ell},K}^{\text{DP}}[\tilde{n}]y_{\hat{\ell},K}^{\text{DP},H}[\tilde{n}-\kappa]$, substituting the value of $y_{\hat{\ell},K}^{\text{DP}}$, the two retained samples are expressed as $r_{y,K}[\kappa]=\sum_{\tilde{n}}y_{\hat{\ell},K}^{\text{DP}}[\tilde{n}D_K]y_{\hat{\ell},K}^{\text{DP},H}[(\tilde{n}-\kappa)D_K]$, and on the original $N$-sample grid, they can be denoted as $\tilde{n}D_K-(\tilde{n}-\kappa)D_K=\kappa D_K$. Through this relation, a lag of one sample in the reduced-rate observation corresponds to the $D_K$ samples at the original sampling rate and is expressed as $\tau=\kappa D_K$. Now the reduced-rate correlation can be visualized on the decimated lag grid as $r_{y,K}[\kappa]\approx r_{y,K}[\kappa D_K]$, and substituting \eqref{filtcorr} with $\lambda$ in $r_{y,K}[\kappa D_K]$ gives
\begin{equation}
    r_{y,K}[\kappa]\approx \sum_{\lambda=-(L_f-1)}^{L_f-1}r_{h_f}[\lambda]r_{u,K}[\kappa D_K-\lambda].
    \label{decicorr}
\end{equation}
Additionally, substituting the channel effects from \eqref{autoco} into \eqref{decicorr} highlights the relation observed in \figurename~\ref{fig_SSdeisng}(c). Aforementioned, in Section~II-C, the mainlobe half-width was defined as $W_{\tau,K}^{(N)}=N/K$, thus, the reduced-rate width corresponds to $W_{\tau,K}^{(M)}=W_{\tau,K}^{(N)}/D_K=M/K$ and is denoted as a scaling factor. Consequently, the mainlobe occupies the same number of samples for $K$-\ac{PRB} configurations at the reduced rate as shown in \figurename~\ref{fig_SSdeisng}(c), and as $D_K$ decreases with the increase in $K$, the temporal correlation resolution still improves with the increasing $K$. For multipath effects at the original $N$ samples, denoting the separation between two channel paths by $\Delta \tau_{a,b}=|\tau_{a}-\tau_{b}|$. Due to decimation, the resulting separation is $\Delta\kappa_{a,b}=|\tau_{a}-\tau_{b}|/D_K$. If $|\tau_{a}-\tau_{b}| < D_K$, then $\Delta \kappa_{a,b} < 1$ results in one sample interval after decimation, which reduces the temporal resolution with which individual multipath components are resolvable and are absorbed in the correlation metric.

\subsection{Narrowband DP-SSB Detection}
After establishing the fact that the \ac{LPSS} correlation structure remains observable after filtering and decimation, the \ac{IoT} receiver must first determine whether the \ac{DPSSB} is present in the M-dimensional observation subspace before initiating \ac{TF} synchronization. At this stage, \ac{DPSSB} symbol boundary knowledge is unavailable; the \ac{IoT} device needs to perform noncoherent detection from the energy received over the candidate observation window. The statistical model of the energy detector is influenced by the effect of the front-end operator on the receiver noise. Before filtering, the \ac{AWGN} samples are independent, whereas the front-end filtering process introduces correlation among the adjacent samples. To highlight this fact, let the filtered noise be denoted by $v_w[n]=\sum_{l=0}^{L_f-1}h_f[l]w_{\ell,K}[n-l]$. Consider the correlation between the two filtered noise samples separated by $\tau$ to be $R_{v_w}[\tau]=\mathbb{E}\{v_w[n]v_w^H[n-\tau]\}$. Substituting $v_w[n]$ in the correlation and since the noise is white, thus, $\mathbb{E}\{v_w[n-l]v_w^H[n-\tau -p]\}=\sigma_w^2\delta[n-l-(n-\tau -p)]$, the correlation outcome depends upon $p=l-\tau$, we get
\begin{equation}
    R_{v_w}[\tau]=\sigma_w^2\sum_{l=0}^{L_f-1}h_f[l]h_f^H[l-\tau].
    \label{noisecorr}
\end{equation}
Substituting the filter autocorrelation in \eqref{noisecorr}, we get $R_{v_w}[\tau]=\sigma_w^2r_{h_f}[\tau]$ which denotes the coloring of noise after filtering. At zero lag, $R_{v_w}[0]=\sigma_w^2\sum_{l=0}^{L_f-1}|h_f[l]|^2$ and filter noise power is defined as $c_K\triangleq\sum_{l=0}^{L_f-1}|h_f[l]|^2$, the variance of the filtered noise sample is transformed to $\sigma_{v,K}^2=c_K\sigma_w^2$. Furthermore, the noise samples are also decimated by $D_K$ as $v_{\ell,K}[\tilde{n}]=v_{w}[\tilde{n}D_K]$. To obtain the noise covariance, consider the two decimated noise samples separated by $\tilde{\kappa}$ samples as 
\begin{equation}
   \begin{aligned}
       R_{v_K}[\tilde{\kappa}]&=\mathbb{E}\{v_K[\tilde{n}]v_K^H[\tilde{n}-\tilde{\kappa}]\},\\
    &=\mathbb{E}\{v_w[\tilde{n}
    D_K]v_w^H[(\tilde{n}-\tilde{\kappa})D_K]\},
    \label{noisecorr1}
   \end{aligned} 
\end{equation}
and the corresponding separation given on the original sampling grid is $R_{v_K}[\tilde{\kappa}]=\sigma_w^2r_{h_f}[\tilde{\kappa}D_K]$. To characterize the effect of the front-end operator on the noise samples, let $r_{h_f}[\tau]=0$ if $|\tau|>L_f-1$ and the decimation factor $D_K>L_f-1$, then the inter-sample $\tilde{\kappa}$ follows the condition $\tilde{\kappa}D_K\geq D_K>L_f-1$, where $R_{v,K}[\tilde{\kappa}]=0,~~\tilde{\kappa}\neq0$, the retained Gaussian samples are uncorrelated and therefore independent. Thus, the covariance output is represented by $\mathbf{C}_{v,K}=c_K\sigma_w^2\mathbf{I}_M$.
  
Following the characterization of the \ac{NB} noise, the \ac{DPSSB} detection must be performed on the \ac{NB} observation produced by the receiver front-end, as given in \eqref{IoTfrontend}, and the detection problem can be expressed through the two hypotheses 
\begin{equation}
    \begin{cases}
\mathcal{H}_0:~\mathbf{y}^{\mathrm{DP}}_{\ell,K}=\mathbf{v}_{\ell,K},\\  \mathcal{H}_1:~\mathbf{y}^{\mathrm{DP}}_{\ell,K}=\boldsymbol{\Phi}_{K}\mathbf{\Omega}_{N}(f_0)\mathbf{H}_{c}\mathbf{P}_{\tau_0}\mathbf{s}_{\ell,K}^{\text{DP}}+\mathbf{v}_{\ell,K},
    \end{cases}
    \label{hypo}
\end{equation}
where $\mathcal{H}_0$ and $\mathcal{H}_1$ represent the absence and presence of the \ac{DPSSB}, respectively.  Consequently, the energy detection statistics can be defined as 
\begin{equation}
\Lambda_K^{\text{ED}}=||\mathbf{y}^{\mathrm{DP}}_{\ell,K}||_{2}^{2}=\mathbf{y}^{\mathrm{DP},H}_{\ell,K}\mathbf{y}^{\mathrm{DP}}_{\ell,K}=\sum_{\tilde{n}=0}^{M-1}|y^{\mathrm{DP}}_{\ell,K}[\tilde{n}]|^2.
\end{equation}
By using the $\Lambda_K^{\text{ED}}$ statistics, the \ac{IoT} device can decide if the \ac{DPSSB} is present or not from the detector $\Lambda_K^{\text{ED}}\mathop{\gtrless}_{\mathcal{H}_0}^{\mathcal{H}_1}\gamma_{d}$, where $\gamma_{d}$ is the detection threshold for the considered $K$ configuration and the desired probability of false alarm as highlighted in Fig.~\ref{fig_61}.  

\subsection{Coarse Timing and CFO Acquisition}
Following the \ac{DPSSB} detection in $\mathcal{H}_1$, the \ac{IoT} receiver must determine the unknown \ac{TO} and \ac{CFO} pairs simultaneously. The objective of this stage is to locate the \ac{DPSSB} on the \ac{NB} timing grid using the known stored \ac{NB} replica of the \ac{LPSS} and estimate $f_0$ coarsely such that the residual frequency offset doesn't interfere with reliable timing acquisition and lies within the unambiguous range of the fine phase-based estimator applied after the \ac{FFT}. Denoting the integer timing shift $\tau_0$ in original-rate samples as $\tau_{0}=D_Kd_0+\epsilon_{\tau,K},$ where $d_0$ represents the \ac{NB} timing shift and $\epsilon_{\tau,K}$ denotes the residual shift within the one \ac{NB} sampling interval. The \ac{CFO} is represented by its phase rotation in the $n$-th original sample rate as $e^{j2\pi f_0n/F_s}$, after decimation, $e^{j2\pi f_0\tilde{n}D_K/F_s}=e^{j2\pi f_0\tilde{n}/F_{s,K}}$. The \ac{CFO} remains unchanged by the decimation; only its normalized value is transformed as $f_{0,K}=\frac{f_0}{F_{s,K}}=D_K\frac{f_0}{F_{s}}.$ Since $F_{s,K}<F_{s}$, the same \ac{CFO} produces a larger phase progression between the adjacent retained samples, which needs to be estimated and compensated to be able to correlate coherently. The correlation process is carried out with a known \ac{NB} \ac{LPSS} that has the same front-end properties incorporated in it and is denoted by 
\begin{equation}
\mathbf{q}_{\hat{\ell},K}=\boldsymbol{\Phi}_{K}\mathbf{s}_{\hat{\ell},K}^\text{ref}=\mathbf{T}_{M}\mathbf{H}_{f}\mathbf{F}_{D,K}\mathbf{d}_{\ell,K}^{\star}.
\end{equation}
Denoting the candidate \ac{NB} timing shift by $d\in \mathcal{T}_K$ and the coarse \ac{CFO} by $f_c\in\mathcal{F_K^C}$ from a predefined search grid. The timing acquisition operating in the decimated region contains the timing hypothesis set defined by $\mathcal{T}_K=\{0,1,\dots,\left\lfloor\tau_{\text{max}}/D_K\right\rfloor\},$ where each $\tau_{\text{max}}$ denotes the original \ac{TO} integer samples. Similarly, let $f_{\text{max}}$ denote the maximum magnitude of the \ac{CFO} candidate and $\Delta f_c$ represent the spacing between the two adjacent \ac{CFO} hypotheses. The coarse \ac{CFO} search set is defined as $\mathcal{F}_K^C=\{-f_{\text{max}}+q\Delta f_c: q=0,1,\dots Q_f-1\},$ where $Q_f=\left\lfloor 2f_{\text{max}}/\Delta f_c\right\rfloor+1$ indicates the number of \ac{CFO} hypotheses. For a candidate $f_c$, \ac{CFO} compensation is expressed as
\begin{equation}
\mathbf{\Omega}_{M}(-f_c)=\mathrm{diag}\{e^{-j2\pi f_c\tilde{n}/F_{s,K}}\}_{\tilde{n}=0}^{M-1}\in\mathbb{C}^{M\times M}.
\end{equation}
The receiver initiates the synchronization procedure by correlating the resulting signal with the shifted reference. The obtained acquisition metric can be expressed as $\mathcal{Z}_{\tau,K}(d)=|\mathbf{q}_{\hat{\ell},K}^H\mathbf{P}_d^{(M,H)}\mathbf{y}_{\ell,K}^{\text{DP}}|^2.$ The \ac{NB} timing estimate is obtained as 
\begin{equation}
\hat{d}_{K}=\arg\max_{d\in\mathcal{T}_K}\{\mathcal{Z}_{\tau,K}(d)\}.
\label{tauest}
\end{equation}
From \eqref{tauest}, the \ac{NB} observation is extracted according to the estimated timing position as $\mathbf{z}_{\ell,K}^{\text{DP}}=\mathbf{S}_{\hat{d}_{K}}\mathbf{y}_{\ell,K}^{\text{DP}}\in\mathbb{C}^{M\times 4}$, where $\mathbf{S}_{\hat{d}_{K}}$ denotes the window selection operator, and the extracted observation is normalized as $\mathbf{\bar{z}}_{\ell,K}^{\text{DP}}=\mathbf{z}_{\ell,K}^{\text{DP}}/||\mathbf{z}_{\ell,K}^{\text{DP}}||_2$ for aiding the coarse \ac{CFO} search. For coarse \ac{CFO} estimation, a \ac{CFO} template bank is constructed, which is then utilized to obtain the coarse \ac{CFO} metric
\begin{equation}
\mathbf{\tilde{y}}_{\ell,K}^{\text{DP}}=\mathbf{\Omega}_{M}(-\hat{f}_{c,K})\mathbf{y}_{\ell,K}^{\text{DP}}.
\end{equation}
The remaining residual frequency offset is $f_{r,K}=f_0-\hat{f}_{c,K}$ with the range of $|f_{r,K}|<f_{\text{amb}}$, where $f_{\text{amb}}$ denotes the unambiguous \ac{CFO} range of the subsequent phase-based fine estimator. To ensure the correct placement of the \ac{FFT} window, the estimated timing shift $\hat{d}_K$ is used to directly position the \ac{NB} observation window. After timing alignment and the coarse \ac{CFO} compensation, the synchronized \ac{TD} received signal is represented by 
\begin{equation}
\mathbf{y}_{\ell,K}^{\text{DP},C}=\mathbf{P}_{-\hat{d}_K}^{(M)}\mathbf{\Omega}_{M}(-\hat{f}_{c,K})\mathbf{y}_{\ell,K}^{\text{DP}}.
\end{equation}
Unlike frequency synchronization, the proposed synchronization design doesn't require a separate fine \ac{TO} estimator. The remaining timing error can be written as $\epsilon_{\tau,K}=\tau_0-D_K\hat{d}_K$, and its sampling grid resolution is bounded by $|\epsilon_{\tau,K}|\leq D_K/2$.

\subsection{Fine CFO Refinement}
The residual \ac{CFO} is further refined before subsequent \ac{FD} processing. The receiver performs an $M$-point \ac{DFT} as $\mathbf{Y}_{\ell,K}^{\text{DP},C}=\mathbf{F}_M^H\mathbf{y}_{\ell,K}^{\text{DP},C}$ to access the known \ac{FD} \ac{LPSS}. By extracting the $\widetilde{\mathcal{D}}_{K}$ known \ac{DMRS} resources as $\mathbf{Z}_{\hat{\ell},K}=\mathbf{S}_{K}^{T,(M)}\mathbf{Y}_{\ell,K}^{\text{DP},C}\in\mathbb{C}^{M\times1}$. For modeling the residual \ac{CFO} between the two \ac{NB} $\hat{\ell}$-th \ac{DPSSB} symbols, it can be expressed as the inter-symbol phase rotation $\Delta \phi_K=2\pi f_{r,K}\Delta N/F_s$, where $\Delta N$ denotes the known temporal separation. Since the dominant effect of the \ac{CFO} is compensated, the residual \ac{CFO} is sufficiently small over the $\hat{\ell}$-th \ac{OFDM} symbols, which can be modeled as a common phase progression and expressed as 
\begin{equation}
\mathbf{Z}_{\hat{\ell},K}=e^{j\phi_{\hat{\ell}}}\mathbf{S}_{K}^{T,(M)}\mathbf{Y}_{\ell,K}^{\text{DP},C}+\mathbf{v}_{\hat{\ell},K}.
\label{freqsymb}
\end{equation}
As the channel is assumed to be approximately constant over the $\hat{\ell}$-th symbols, the \ac{FD} inner product is expressed as $\mathbf{Z}_{\hat{\ell}_1,K}^H\mathbf{Z}_{\hat{\ell}_3,K}$. The symbols can exhibit deterministic phase differences; to compensate for this effect, the two-symbol cross-product is calibrated by a reference cross-product as
$\mathcal{\zeta}_K\triangleq\sum_{k\in\widetilde{\mathcal{D}}_K}(Z_{\hat{\ell}_1,K}^*[\bar{k}]Z_{\hat{\ell}_3,K}[\bar{k}])(Z_{\hat{\ell}_1,K}^{*,\mathrm{ref}}[\bar{k}]Z_{\hat{\ell}_3,K}^{\mathrm{ref}}[\bar{k}])^*$, where $\bar{k}$ represents the narrowband subcarrier index corresponding to $k\in\widetilde{\mathcal{D}}_K$.

Ignoring noise for now, the final expression is $\mathcal{\zeta}_K=e^{j\Delta \phi_{K}}\sum_{k\in\widetilde{\mathcal{D}}_K}|(Z_{\hat{\ell}_1,K}^*[\bar{k}]Z_{\hat{\ell}_3,K}[\bar{k}])(Z_{\hat{\ell}_1,K}^{*,\mathrm{ref}}[\bar{k}]Z_{\hat{\ell}_3,K}^{\mathrm{ref}}[\bar{k}])^*|$, and as $\sum_{k\in\widetilde{\mathcal{D}}_K}|(Z_{\hat{\ell}_1,K}^*[\bar{k}]Z_{\hat{\ell}_3,K}[\bar{k}])(Z_{\hat{\ell}_1,K}^{*,\mathrm{ref}}[\bar{k}]Z_{\hat{\ell}_3,K}^{\mathrm{ref}}[\bar{k}])^*|\in\mathbb{R}\geq0$, no phase is contributed; thus, the channel-independent residual \ac{CFO} is measured as $\angle \mathcal{\zeta}_K=\Delta \phi_{K}=2\pi f_{r,K}\Delta N/F_s (\mathrm{mod}~2\pi)$. Denoting $\widehat{\Delta \phi_K}=\angle(\mathcal{\zeta}_K),$ with $\widehat{\Delta \phi_K}\in(-\pi,\pi]$, the \ac{CFO} estimate is written as 
\begin{equation}
\hat{f}_{r,K}=\frac{F_s}{2\pi \Delta N}\angle(\mathcal{\zeta}_K).
\label{freqsymb1}
\end{equation}
The total \ac{CFO} estimate is given by $\hat{f}_{0,K}=\hat{f}_{c,K}+\hat{f}_{r,K}$. The unambiguous range of the fine estimator needs to be characterized as $-\pi <\angle \zeta_K\leq\pi$, which leads to $-\frac{F_s}{2\Delta N}<f_{r,K}\leq\frac{F_s}{2 \Delta N}$. Defining the maximum fine \ac{CFO} estimate range as $f_{\mathrm{fine,max}}\triangleq\frac{F_s}{2 \Delta N}$, the phase-based estimator unambiguous region is $|f_{r,K}|<f_{\mathrm{fine,max}}$. Consequently, the coarse estimator must also ensure $|f_{0}-\hat{f}_{c,K}|<\frac{F_s}{2 \Delta N}$.

\section{Performance Analysis}
This section presents the theoretical analysis of the proposed \ac{DPSSB} framework for different performance metrics such as detection probability, synchronization error, cost of complexity, and \ac{BER}.

\subsection{DP-SSB Detection Performance}
Following the binary hypotheses developed in \eqref{hypo}, for an $M$-sample detection observation, under $\mathcal{H}_0$, the normalized energy statistics follow a central chi-square distribution as presented in \cite{1204119}, and can be expressed as $\left.\frac{2\Lambda_K^{\mathrm{ED}}}
{\sigma_{v,K}^2}\right|\mathcal{H}_0\sim\chi_{2M}^{2}$. The conditional \ac{PDF} of $\Lambda_K^{\mathrm{ED}}$ is expressed as
\begin{equation}
f_{\Lambda_K^{\mathrm{ED}}|\mathcal{H}_0}(x)=\frac{x^{M-1}}{\Gamma(M)(\sigma_{v,K}^2)^M}\exp(-\frac{x}{\sigma_{v,K}^2}),~x\geq0,
\end{equation}
where $\Gamma(\cdot)$ denotes the Gamma function. Denoting the probability of false alarm for the decision rule $\Lambda_K^{\text{ED}}\mathop{\gtrless}_{\mathcal{H}_0}^{\mathcal{H}_1}\gamma_{d}$ as
\begin{equation}
P_{\mathrm{FA}}=1-F_{\chi_{2M}^2}\big(\frac{2\gamma_d}{\sigma_{v,K}^2}\big),
\end{equation}
where $F_{\chi_{2M}^2}(.)$ denotes the \ac{CDF} of a central chi-square random variable with $2M$ degrees of freedom. Therefore, for a desired $P_{\mathrm{FA}}$, the required detection threshold is 
\begin{equation}
\gamma_d=\frac{\sigma_{v,K}^2}{2}F_{\chi_{2M}^2}^{-1}(1-P_{\mathrm{FA}}).
\label{single_symbol_fixed_threshold}
\end{equation}
Under $\mathcal{H}_1$, the received observation includes $\boldsymbol{\Phi}_{K}\mathbf{\Omega}_{N}(f_0)\mathbf{H}_{c}\mathbf{P}_{\tau_0}\mathbf{s}_{\ell,K}^{\text{DP}}$ in addition to $\mathbf{v}_{\ell,K}$. Consequently, the normalized energy statistics follow a non-central chi-square distribution expressed as $\left.\frac{2\Lambda_K^{\mathrm{ED}}}
{\sigma_{v,K}^2}\right|\mathcal{H}_1\sim\chi_{2M}^{'2}(\lambda_K)$, where $\lambda_K=\frac{2||\boldsymbol{\Phi}_{K}\mathbf{\Omega}_{N}(f_0)\mathbf{H}_{c}\mathbf{P}_{\tau_0}\mathbf{s}_{\ell,K}^{\text{DP}}||_2^2}{\sigma_{v,K}^2}$, and denotes the non centrality parameter. The resulting \ac{PDF} is given as
\begin{equation}
\begin{aligned}
f_{\Lambda_K^{\mathrm{ED}}|\mathcal{H}_1}(x)&=\frac{1}{\sigma_{v,K}^2}\exp{\left(-\frac{x}{\sigma_{v,K}^2}-\frac{\lambda_K}{2}\right)}\left(\frac{2x}{\sigma_{v,K}^2\lambda_K}\right)^{\frac{M-1}{2}}\\& \times I_{M-1}\left(\sqrt{\frac{2\lambda_Kx}{\sigma_{v,K}^2}}\right),~x\geq0,
\end{aligned}
\end{equation}
where $I_{M-1}(.)$ denotes the modified Bessel function of the first kind of order $M-1$. The corresponding probability of correctly detecting the \ac{DPSSB} is
\begin{equation}
    P_{D,K}=1-F_{\chi_{2M}^{'2}(\lambda_K)}\left(\frac{2\gamma_d}{\sigma_{v,K}^2}\right)=Q_M\left(\sqrt{\lambda_K},\sqrt{\frac{2\gamma_d}{\sigma_{v,K}^2}}\right),
    \label{pdmq}
\end{equation}
where $Q_{M}\left(\cdot,\cdot\right)$ represents the generalized Marcum-$Q$ function expressed in its closed form.

\subsection{Synchronization Estimation Performance}
The synchronization performance of the proposed \ac{IoT} receiver is evaluated in terms of estimation errors linked to \ac{NB} timing acquisition, coarse, and phase-based fine \ac{CFO} refinement.  
\cite{650240, 599949, 328961}

\subsubsection{Timing Error Metric}
Defining the timing error as $e_{\tau.K}=\hat{\tau}_K-\tau_0$. Since the timing acquisition is performed after the decimation, the estimator is reduced to the \ac{NB} timing grid, which is $\hat{\tau}_K=D_K\hat{d}_K$. The achievable timing resolution is determined by the reduced sampling interval. Since the timing error is dominated by noise, multipath, and uncompensated \ac{CFO}, the timing estimation error is evaluated using its \ac{RMSE}, and is defined as $\mathrm{RMSE}_{\tau,K}=\sqrt{\frac{1}{N_{tr}}\sum_{i=1}^{N_{tr}}|\hat{\tau}_K^{(i)}-\tau_0|^2}$. 

\subsubsection{CFO Error Metrics}
 Following the timing acquisition, the coarse \ac{CFO} estimation error is given by $e_{f,K}^C=\hat{f}_{c,K}-f_0$. Each adjacent \ac{CFO} hypotheses on the grid are separated by $\Delta f_c$, the correct selected hypothesis results in the grid limited error given by $|e_{f.K}^{C,\mathrm{grid}}|\leq \Delta f_c/2$, besides this, the errors caused by the noise, multipath, and timing errors can be characterized by $\mathrm{RMSE}_{f,K}^C=\sqrt{\frac{1}{N_{tr}}\sum_{i=1}^{N_{tr}}|\hat{f}_{c,K}^{(i)}-f_0|^2}$. An additional requirement of the \ac{CFO} estimator is to place the residual \ac{CFO}, which is $f_{r,K}=f_0-\hat{f}_{c,K}$, within the unambiguous range of the phase-based fine estimator. The final \ac{CFO} results in the final \ac{CFO} estimator error being defined as $e_{f,K}^F=\hat{f}_{0,K}-f_0$, and its performance is characterized by the \ac{RMSE} as $\mathrm{RMSE}_{f,K}^F=\sqrt{\frac{1}{N_{tr}}\sum_{i=1}^{N_{tr}}|\hat{f}_{0,K}^{(i)}-f_0|^2}$.

\subsection{Computational Complexity Analysis}
The computational complexity of the proposed synchronization scheme is evaluated in terms of \ac{CM}s and compared with SC, ML, and LoRa synchronization schemes \cite{650240, 599949, 9000820}. A \ac{CM} and a magnitude-square operation are each counted as one \ac{CM}, while additions, comparisons, indexing, and phase extractions are neglected. An $M$-point radix-$2$ \ac{FFT} requires $\mathcal{O}(\frac{M}{2} \mathrm{log}_2 M)$ \ac{CM}s, the comparison is restricted to synchronization only; thus, filtering, decimation, and energy detection operations are excluded. For the proposed scheme, the timing acquisition searches along the candidate set $\mathcal{T}_K$ and the correlation with the known reference requires $M$ \ac{CM}s, resulting in $\mathcal{O}_{\mathrm{TA},K}=\mathcal{O}(|\mathcal{T}_K|M)$. After timing synchronization, frequency synchronization includes a coarse \ac{CFO} estimator that evaluates the correlation metric over the predefined frequency-hypothesis set $\mathcal{F}_K^C$, requiring $M$ inner products, which leaves the cost of complexity to be $\mathcal{O}_{\mathrm{CFO},K}^C=\mathcal{O}(|\mathcal{F}_K^C|M)$, and is compensated over the whole received signal observation, earning up to $4M$ \ac{CM}s. Following the coarse \ac{CFO} compensation, the $\hat{\ell}$-th \ac{DPSSB} symbols are transformed to \ac{FD} using two $M$-point \ac{FFT} with complexity $\mathcal{O}(M\mathrm{log}_2 M)$. After that, the residual \ac{CFO} is estimated via the phase difference between the $\hat{\ell}$-th \ac{DPSSB} symbols using an element-wise product that was calibrated with stored references via cross product, incurring an additional $2M$ \ac{CM}s. Consequently, the overall synchronization complexity of the proposed \ac{IoT} receiver is
\begin{equation}
\mathcal{O}_{\mathrm{Prop},K}=\mathcal{O}(M(|\mathcal{T}_K|+|\mathcal{F}_K^C|+6+\mathrm{log}_2 M)).
\end{equation}
\begin{table}[!t]
\caption{Simulation Parameters\label{tab:table1}}
\centering
\begin{tabular}{|c|c|}
\hline
\textbf{Parameters} & \textbf{Values}\\
\hline
Number of carriers $N_c$ & $240$\\
\hline
Selected \ac{FFT} symbols $K$ & $\{12, 24, 48\}$\\
\hline
\ac{FFT} size $N$ & $256$\\
\hline
Carrier frequency $f_c$ & $2.4$ GHz\\
\hline
Subcarrier spacing $\Delta f$ & $15$ kHz\\
\hline
\ac{IoT} bandwidth & $\{180, 360, 720\}$ kHz\\
\hline
\ac{SSB} bandwidth & $3.6$ MHz\\
\hline
\ac{CP} size & $N/8$\\
\hline
Modulation order & $4$-QAM\\
\hline
Channel model & \ac{3GPP} \ac{TDL}-C\\
\hline
Probability of false alarm & $P_{\mathrm{FA}} = \{10^{-2}, 10^{-4}\}$\\
\hline
\ac{CFO} & $\epsilon = 5$ kHz\\
\hline
Monte Carlo trials & $N_{mc} = 10^{5}$\\
\hline
\end{tabular}
\end{table}
\subsection{NR UE Communication Performance}
To highlight the effect of the modified \ac{DMRS} relocation at the \ac{NR} \ac{UE} receiver during channel estimation and the \ac{PBCH} data recovery, perfect \ac{TF} synchronization is assumed as given in \cite{tuninato2023comprehensive}. Furthermore, the \ac{UE} is assumed to know both the modified \ac{DMRS} locations and the corresponding \ac{LS}-shaped coefficients $\mathbf{d}_{\ell,K}^{\star}$ defined earlier in Section~II-C. After \ac{CP} removal and the conventional $N$-point \ac{FFT}, the received \ac{DPSSB} signal is $Y_{K}^{\mathrm{DP}}[k,\ell]
=X_{K}^{\mathrm{DP}}[k,\ell]H[k]
+W[k,\ell],$ where $H[k]\triangleq\sum_{p=0}^{P_h-1}h_pe^{-j\frac{2\pi}{N}k\tau_p}$ and $W[k,\ell]$ denotes the \ac{FD} noise. Under perfect \ac{CSI} conditions, the reference \ac{PBCH} performance is obtained as the lower bound. For the imperfect-\ac{CSI} case, all \ac{DMRS} allocations are utilized for channel estimation. In contrast to the conventional \ac{SSB}, the \ac{DPSSB} contains both the remaining distributed \ac{DMRS} and the modified \ac{DMRS} block over $\widetilde{\mathcal{D}}_K$. Hence, the resulting channel estimation depends on both the \ac{DMRS} locations and their transmitted coefficients \cite{1033876}. For this purpose, the received signal model can be written as
\begin{equation}
\mathbf{y}_{\mathrm{D},K}=\mathbf{U}_{\mathrm{D},K}\mathbf{h}_{c}+\mathbf{w}_{\mathrm{D},K} \in \mathbb{C}^{N\times 1},
\label{eq:ue_dmrs_observation}
\end{equation}
where $\mathbf{w}_{\mathrm{D},K}$ denotes the noise and each row of $\mathbf{U}_{\mathrm{D},K}$ corresponds to  \ac{DMRS} symbols as $[\mathbf{U}_{\mathrm{D},K}]_{k,p}
=X_{K}^{\mathrm{DP}}[k,\ell]e^{-j\frac{2\pi}{N}kp}$. Denoting the delay-domain \ac{LS} channel estimation given in \cite{504981}, by $\widehat{\mathbf{h}}_{K}^{\mathrm{LS}}
=\mathbf{U}_{\mathrm{D},K}^{\dagger}\mathbf{y}_{\mathrm{D},K},$ For an achievable full-column-rank $\mathbf{U}_{\mathrm{D},K}$, the corresponding estimation-error covariance is given by
\begin{equation}
\mathbb{E}\left[\left(\widehat{\mathbf{h}}_{K}^{\mathrm{LS}}-\mathbf{h}_{c}\right)\left(\widehat{\mathbf{h}}_{K}^{\mathrm{LS}}-\mathbf{h}_{c}\right)^H\right]=\sigma_w^{2}(\mathbf{U}_{\mathrm{D},K}^H\mathbf{U}_{\mathrm{D},K})^{-1}.
\label{eq:ue_ls_error_covariance}
\end{equation}
From~\eqref{eq:ue_ls_error_covariance}, it can be observed that although the total number of \ac{DMRS} symbols is preserved, their relocation and \ac{LS} optimization alter $\mathbf{U}_{\mathrm{D},K}^H\mathbf{U}_{\mathrm{D},K}$, and can therefore affect the channel-estimation accuracy. 

To further highlight the performance trade-off, a \ac{LMMSE} channel estimator presented in \cite{504981} is considered. Let the channel covariance matrix be denoted by $\mathbf{R}_{h_c}=\mathbb{E}[\mathbf{h}_c\mathbf{h}_c^H]$, and the estimator is defined as
\begin{equation}
\widehat{\mathbf{h}}_{K}^{\mathrm{LMMSE}}=\mathbf{R}_{h_c}\mathbf{U}_{\mathrm{D},K}^H(\mathbf{U}_{\mathrm{D},K}\mathbf{R}_{h_c}\mathbf{U}_{\mathrm{D},K}^H+\sigma_w^{2}\mathbf{I})^{-1}\mathbf{y}_{\mathrm{D},K}.
\label{LMMSE}
\end{equation}
The estimator in \eqref{LMMSE} exploits both the known \ac{DPSSB} \ac{DMRS} structure and the second-order channel statistics. The \ac{CFR} over the $N_c$ \ac{SSB} subcarriers is reconstructed for both estimators as $\widehat{H}_K[k]=\sum_{p=0}^{L_h-1}\widehat{h}_K[p]e^{-j\frac{2\pi}{N}kp}$ and the \ac{PBCH} symbols are equalized as $\widehat{X}_{K}^{\mathrm{PBCH}}[k,\ell]
=\frac{\widehat{H}_K^{*}[k]}{|\widehat {H}_K[k]|^2+\widehat{\sigma}_w^2}Y_{K}^{\mathrm{DP}}[k,\ell],$ where $\widehat{\sigma}_w^2$ denotes the noise variance used by the equalizer. For the perfect-\ac{CSI} and \ac{LMMSE} cases, $\widehat{\sigma}_w^2=\sigma_w^2$, whereas for delay-domain \ac{LS} it is obtained from the \ac{DMRS} residual. Finally, the coded \ac{PBCH} \ac{BER} is evaluated as $\mathrm{BER}=\frac{1}{N_{\mathrm{mc}}N_b}
\sum_{i=1}^{N_{\mathrm{mc}}}\sum_{n=1}^{N_b}
\mathbbm{1}\left\{b_{i,n}\neq\widehat b_{i,n}
\right\},$ where $N_b$ denotes the number of \ac{PBCH}coded bits. The perfect-\ac{CSI} isolates the \ac{PBCH} detection performance from the channel-based errors, whereas the \ac{LS} and \ac{LMMSE} cases characterize the communication penalty incremented from the modified \ac{DPSSB} \ac{DMRS} geometry and coefficients.

\section{Simulation Results}
In this section, the performance of the proposed \ac{DPSSB} synchronization framework is assessed through numerical simulations under different $K$ and $P_\mathrm{FA}$ values for probability of detection, synchronization \ac{RMSE}, computational complexity, and \ac{BER}. The simulation parameters utilized in the evaluation of the proposed techniques are summarized in Table \ref{tab:table1}. The specific normalized CFO value is chosen based on the \ac{IoT} crystal oscillator quality as mentioned in \cite{8032502}.
\begin{figure}[!t]
\centering
\includegraphics[width=2.5in]{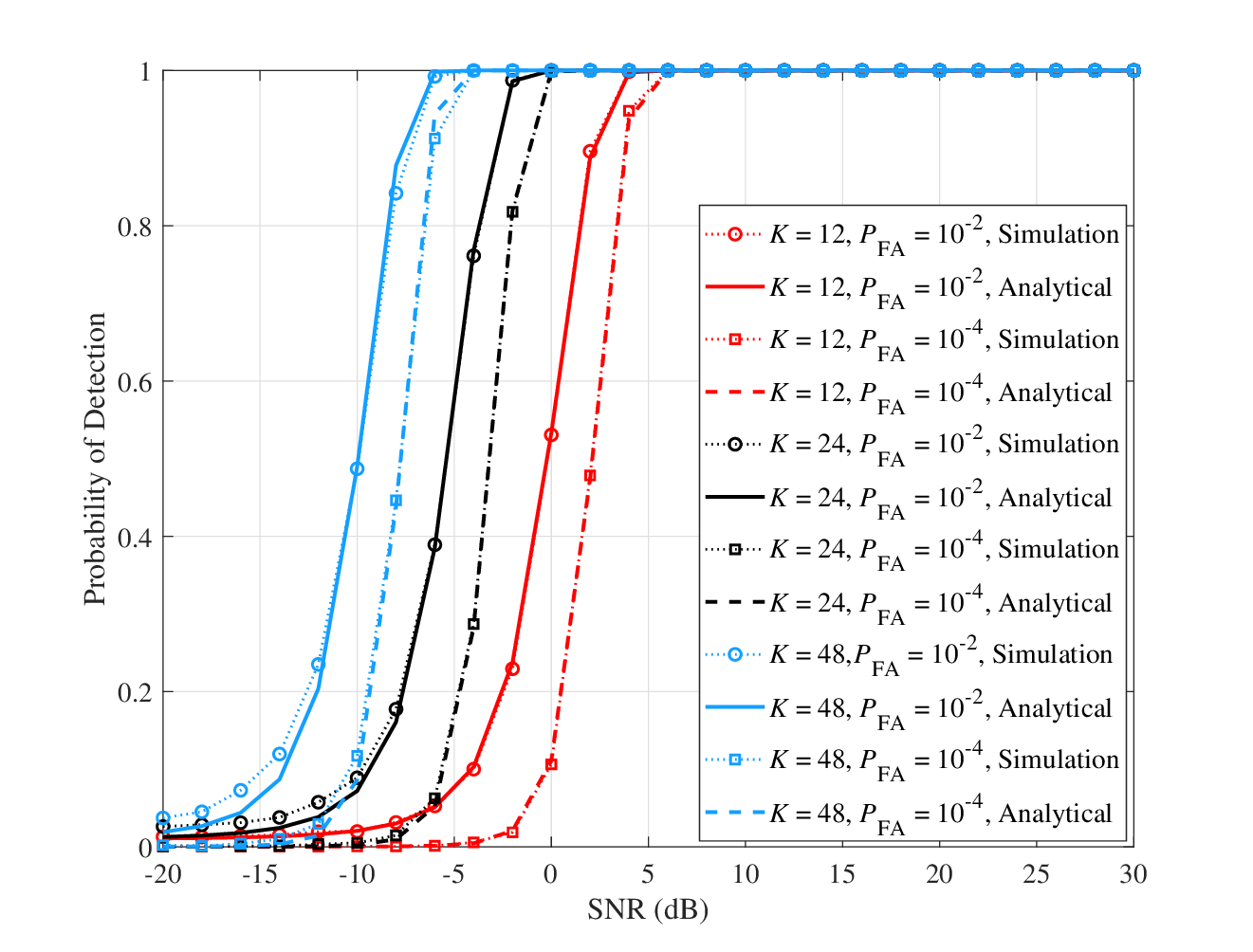}
\caption{Probability of detection versus \ac{SNR} for different $K$ and $P_{\mathrm{FA}}$ values.}
\label{fig_PD}
\end{figure}

In \figurename~\ref{fig_PD}, the probability of detection as a function of \ac{SNR} is plotted for different $K$ and $P_{\mathrm{FA}}$ values. As illustrated in \figurename~\ref{fig_PD}, the analytical results closely follow the Monte Carlo simulations over the \ac{SNR} range, validating the derived detection expression \eqref{pdmq}. The detection probability increases monotonically for all $K$ with \ac{SNR} and approaches unity as the receiver is able to distinguish the received \ac{LPSS} from the noise in the extremely low \ac{SNR} region. As $K$ is increased, the detection region is shifted towards the low \ac{SNR} regime; with $K=48$ the best detection performance is obtained, followed by $K=24$ and $K=12$, which is due to the increased observation energy determined by the detector. Furthermore, reducing the $P_{\mathrm{FA}}$ from $10^{-2}$ to $10^{-4}$ raises the detection threshold and therefore requires a higher \ac{SNR} to achieve the same detection performance.

\begin{figure}[!t]
\centering
\includegraphics[width=\linewidth]{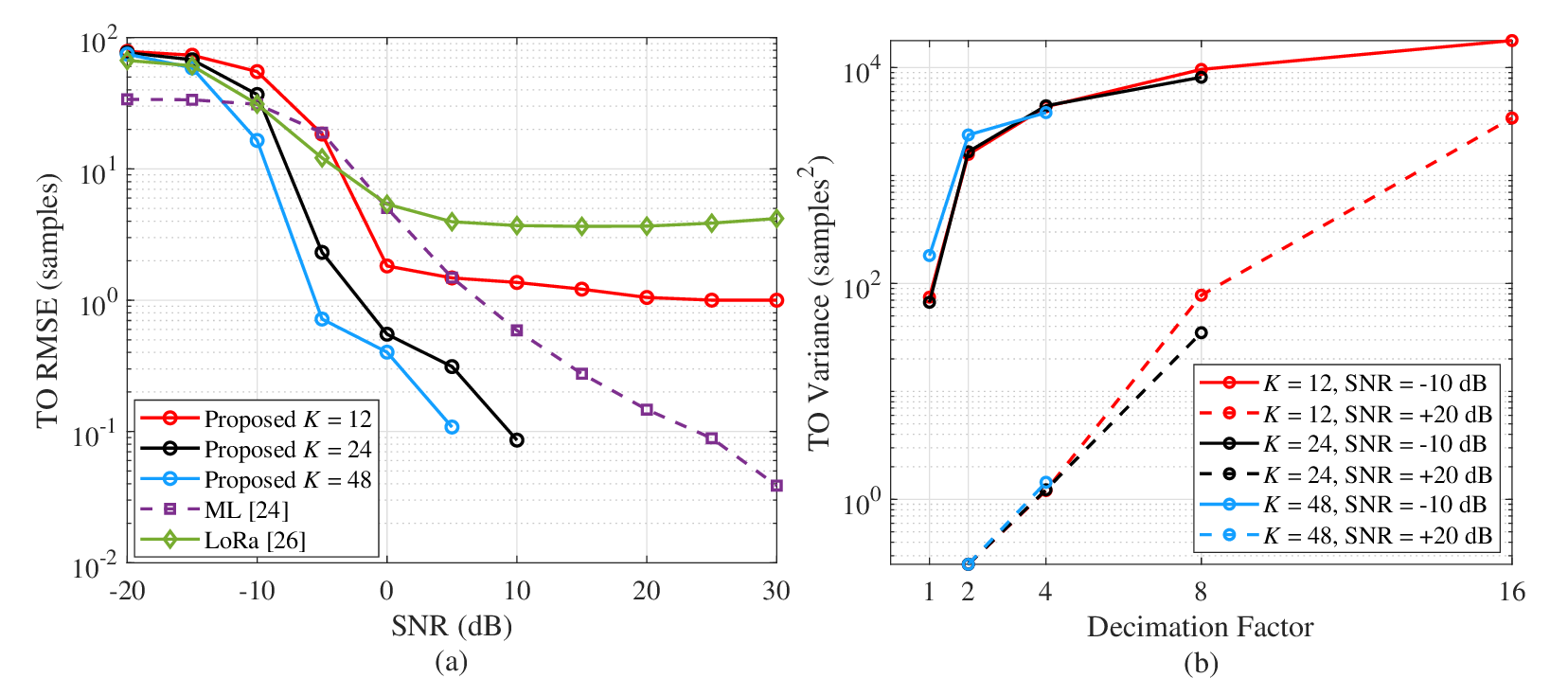}
\caption{\ac{TO} performance: (a) \ac{TO} \ac{RMSE} versus \ac{SNR} for $K$ values ; (b) \ac{TO} variance versus decimation factor for $K$ values under \ac{SNR} of $-10$ dB and $20$ dB.}
\label{fig_timingRMSE}
\end{figure}

\figurename~\ref{fig_timingRMSE}~(a) depicts the \ac{TO} \ac{RMSE} as a function of \ac{SNR} for $K$ values, which is compared with the baseline synchronization schemes including ML \cite{599949} and LoRa \cite{9000820}. The timing acquisition performance of the proposed receiver is clearly depicted to be $K$-dependent, and as $K$ increases, the timing accuracy increases substantially. In particular, $K=48$ achieves the lowest \ac{TO} \ac{RMSE}; in comparison, smaller $K$ values illustrate higher \ac{RMSE} performance. For $K=12$, the proposed \ac{TO} \ac{RMSE} encounters an error floor due to the restricted temporal resolution, which is a consequence of decimation. This behavior directly conforms with the \ac{LPSS} bandwidth and the shrinking of the correlation structure, which results in increased timing resolution. At high \ac{SNR} values, the proposed \ac{LPSS} curves for $K=24$ and $K=48$ perform better than the benchmark schemes. Furthermore, the effect of receiver decimation on timing resolution is investigated in \figurename~\ref{fig_timingRMSE}~(b). \ac{TO} variance as a function of decimation factor is illustrated in \figurename~\ref{fig_timingRMSE}~(b) at \ac{SNR} of $-10$ dB and $20$ dB. As the decimation factor is increased, the variance becomes larger; for $K=12$, the variance is the highest because fewer timing samples are available, and the temporal resolution is the worst. Compared with $K=24$, the number of timing samples increases, resulting in a reduction in variance, which increases the resolution, and for $K=48$ the timing resolution is the highest, as more samples are available after decimation, which further reduces the variance.

 \begin{figure}[!t]
\centering
\includegraphics[width=1\linewidth]{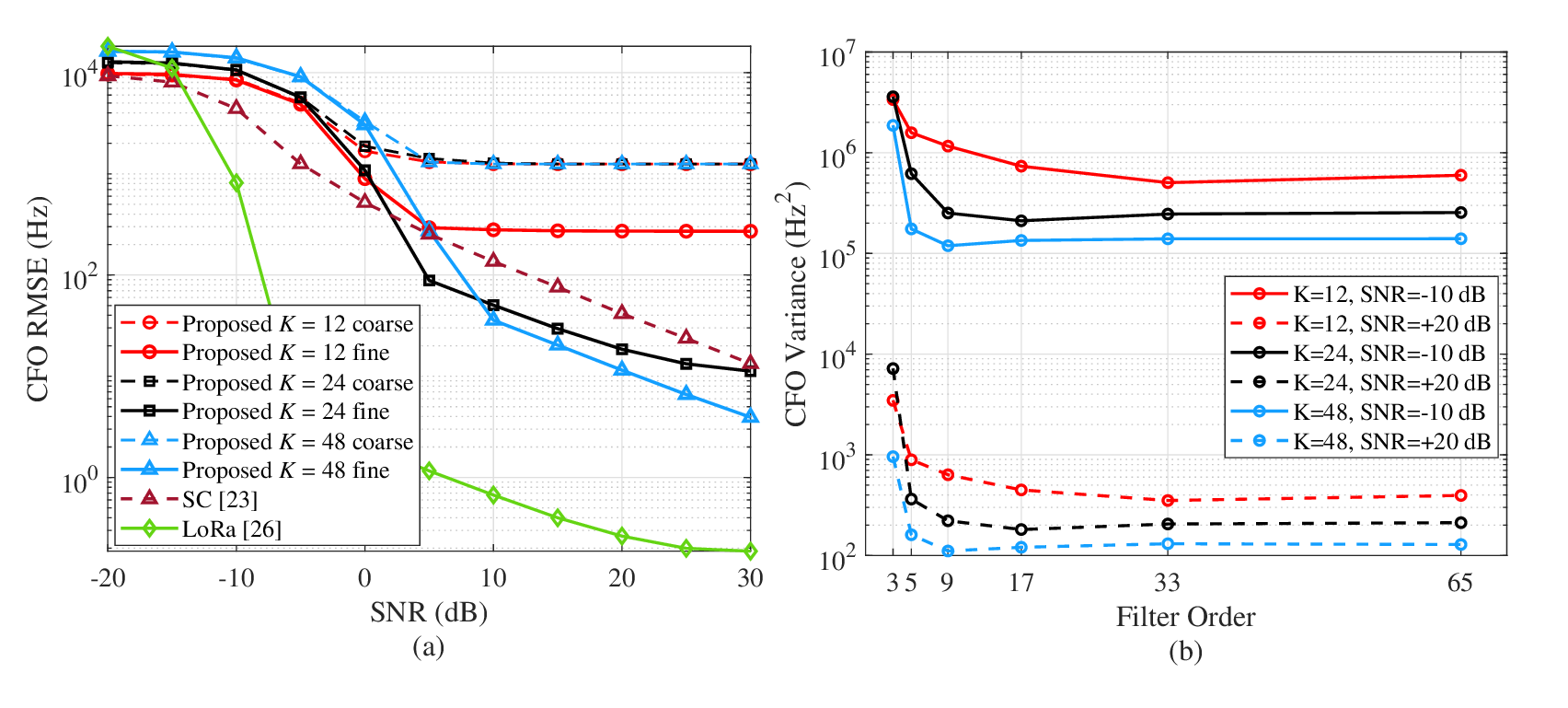}
\caption{\ac{CFO} Performance: (a) coarse and fine \ac{CFO} \ac{RMSE} versus \ac{SNR} for $K$ values; (b) \ac{CFO} variance versus filter order for $K$ values under \ac{SNR} of $-10$ dB and $20$ dB.}
\label{fig_cfo}
\end{figure}

In \figurename~\ref{fig_cfo}~(a), the \ac{CFO} \ac{RMSE} of the coarse and fine estimators is depicted as a function of \ac{SNR} for different $K$ values and compared against baseline methods such as SC \cite{650240} and LoRa \cite{9000820}. The coarse estimates exhibit large errors for smaller $K$ values and approach $K$-dependent floors at moderate \ac{SNR} values. The fine estimates exhibit lower errors for all $K$ values with increasing \ac{SNR}, and particularly, $K=48$ provides the lowest final \ac{CFO} \ac{RMSE} among the $K$ values reaching the Hz-level at high \ac{SNR} values. The difference in performance conveys the operational limits of both estimators, as coarse resolves the initial \ac{CFO} ambiguity, and the fine estimator provides residual \ac{CFO} refinement. Comparing the reduced-rate proposed estimators with the baseline methods, SC provides a competitive \ac{CFO} \ac{RMSE} accuracy as \ac{SNR} increases but operates on the full sampling rate. LoRa provides a separate low-power synchronization reference and achieves the lowest \ac{RMSE} due to the repetitive structure of preambles available for synchronization. \figurename~\ref{fig_cfo}~(b) illustrates the effect of the receiver filter on the sensitivity of the \ac{CFO} estimators. \ac{CFO} variance as a function of filter order is presented in \figurename~\ref{fig_cfo}~(b) for $K$ values under \ac{SNR} of $-10$ dB and $20$ dB. Increasing the filter order reduces the \ac{CFO} variance as the receiver obtains a better isolated band. After filter order $33$, the additional filter taps provide saturated performance. The variance of \ac{CFO} is particularly lower at $20$ dB than at $-10$ dB for all $K$. The variance of $K=48$ is the lowest among the $K$ values, as more frequency bins are available for the estimators. 
\begin{figure}[!t]
\centering
\includegraphics[width=\linewidth]{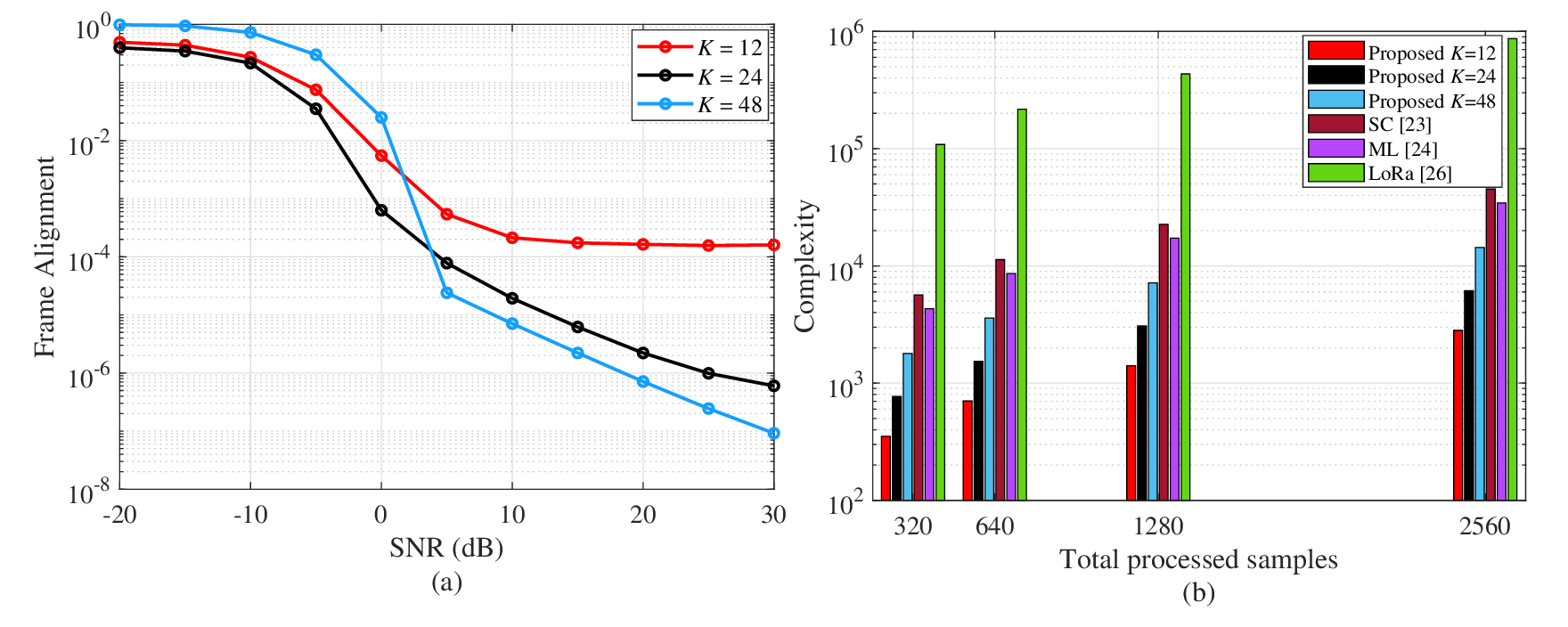}
\caption{\ac{IoT} receiver performance: (a) Effective frame alignment versus \ac{SNR} for $K$ values; (b) System's computational complexity as a function of the total processed samples for $K$ values.}
\label{fig_FA}
\end{figure}
 \begin{figure}[!t]
\centering
\includegraphics[width=\linewidth]{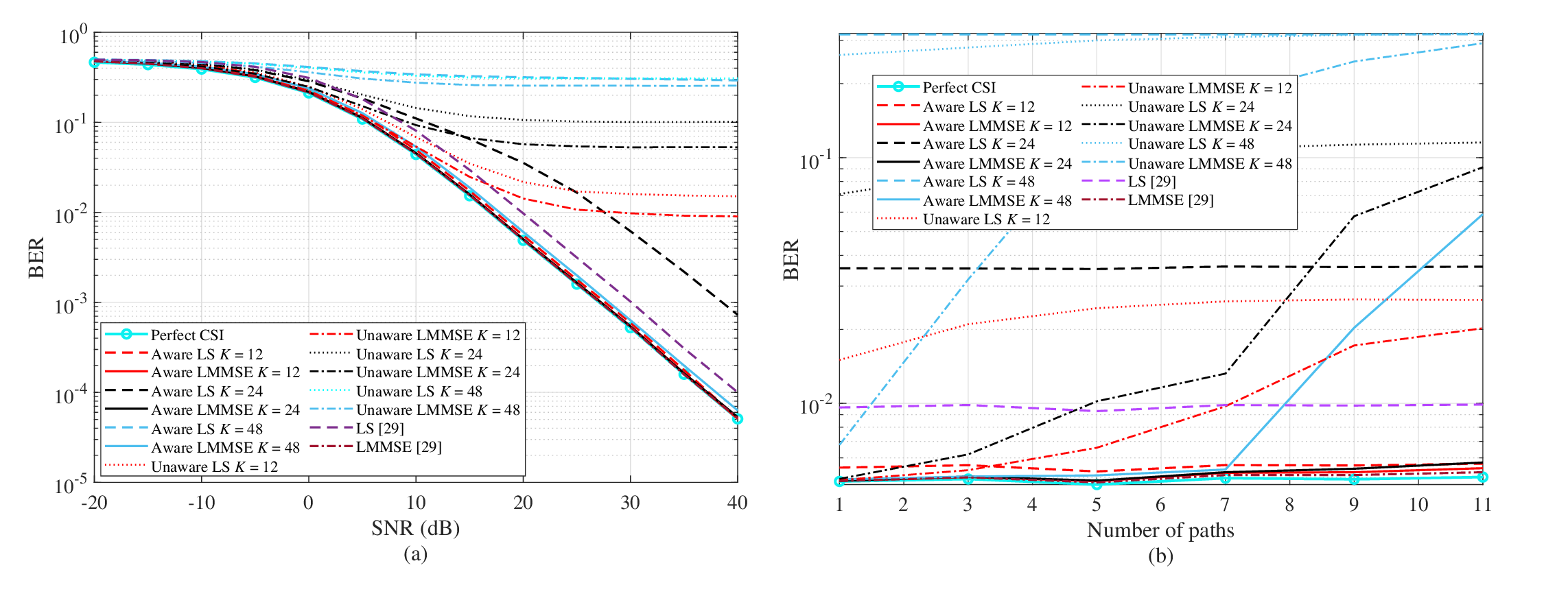}
\caption{\ac{BER} performance: (a) \ac{BER} versus \ac{SNR} of $K$ values under \ac{DMRS} aware and un-aware conditions; (b) \ac{BER} versus number of taps for a fixed $10$ dB \ac{SNR} with $K$ values under \ac{DMRS} aware and un-aware conditions.}
\label{fig_BERtaps}
\end{figure}

\figurename~\ref{fig_FA} presents the overall synchronization reliability and \ac{IoT} receiver complexity. In \figurename~\ref{fig_FA}~(a), the effective frame alignment as a function of \ac{SNR} for $K$ values is illustrated. The frame alignment metric jointly accounts for \ac{TO} and \ac{CFO} variances to highlight synchronization performance. As $K$ increases, the frame alignment error decreases with increasing \ac{SNR}. For $K=48$, more timing and frequency synchronization information is available as a consequence of larger bandwidth and results in the lowest frame alignment error, followed by $K=24$ and $K=12$. The highest frame alignment error is encountered by $K=12$ due to severely restricted timing and frequency resolution. As shown in \figurename~\ref{fig_FA}~(b), the synchronization complexity as a function of total processed samples is plotted and measured in \ac{CM}s for $K$ values and baseline methods including SC \cite{650240}, ML \cite{599949}, and LoRa \cite{9000820}. For all schemes, the computational cost increases with the increase in the observation length, and the rate of increase depends upon the synchronization procedure. As seen from \figurename~\ref{fig_FA}~(b), $K=2$ incurs the lowest complexity cost among all proposed and baseline synchronization schemes, reflecting the timing and frequency synchronization operations in the reduced $M$ observational dimension. Alternatively, conventional SC and ML methods sustain a higher complexity cost because of operating in the full-rate observation regime. Moreover, LoRa exhibits the highest complexity cost due to the repeated observation lengths under the adopted implementation \cite{9000820}. 

\figurename~\ref{fig_BERtaps}~(a) presents the \ac{PBCH} \ac{BER} performance as a function of \ac{SNR} under perfect \ac{CSI} reference for $K$ values. Under \ac{DMRS}-aware \ac{LS} and \ac{LMMSE} channel estimation, \ac{BER} consistently decreases and approaches the perfect \ac{CSI} reference \ac{BER}, demonstrating that the relocated and modified \ac{DMRS} can still support reliable channel estimation when the modified \ac{DMRS} is known to the \ac{UE}. The \ac{LMMSE} estimator provides the strongest robustness due to exploiting the channel covariance information, and the \ac{DMRS}-unaware estimators exhibit pronounced \ac{BER} floors. As $K$ increases, the \ac{DMRS} structure-aware \ac{LS} estimator earns an increasing error floor compared with the \ac{LMMSE}. In the case of \ac{DMRS}-unaware, as $K$ increases, both the \ac{LS} and \ac{LMMSE} estimators experience increasing error floors due to mismatch in the \ac{DMRS} structure. Alternatively, \ac{BER} is further assessed in terms of the number of channel taps as shown in \figurename~\ref{fig_BERtaps}~(b). Under perfect \ac{CSI}, the \ac{BER} remains insensitive to the channel taps, providing a reference that acts as a baseline for the effect of channel estimation error. The \ac{DMRS}-aware \ac{LMMSE} estimator provides the best \ac{BER} performance even at higher channel taps, compared to the \ac{LS} estimator, whose BER performance degrades with the increase in $K$. In contrast, for the \ac{DMRS}-unaware \ac{LS} and \ac{LMMSE} estimators, the BER performance decreases with the increase in $K$ and as the number of channel taps increases. At sufficient channel length, both estimators' \ac{BER} performance is almost equal. This performance degradation exposes the tradeoff between increasing synchronization accuracy with increasing $K$ and preserving \ac{DMRS} structural geometry for the \ac{PBCH} reception.

\section{Conclusion}
In this work, a novel \ac{DPSSB} design is proposed to support low-power \ac{IoT} synchronization, while preserving conventional \ac{UE} operation within the same \ac{IA} signaling block. In the proposed design, selected \ac{DMRS} \ac{RE}s are relocated into a single or multiple \ac{PRB}s, followed by \ac{LS}-optimized to obtain a sharp \ac{TD} signal shape after filtering and decimation. This desirable correlation structure is leveraged to perform energy-based detection, timing acquisition, and coarse \ac{CFO} estimation in the reduced-dimensional subspace, followed by \ac{FD} phase-based fine \ac{CFO} synchronization using the two modified \ac{DPSSB} symbols. The performance of the proposed design is evaluated in terms of detection probability, \ac{TF} synchronization \ac{RMSE}, computational complexity, and \ac{UE} side \ac{PBCH} performance under perfect-\ac{CSI}, \ac{LS}, and \ac{LMMSE} channel estimation. Simulation results validate the analytical detection characterization, demonstrate reliable \ac{IoT} \ac{TF} synchronization, and illustrate meaningful \ac{UE} communication performance. These results demonstrate the feasibility of a unified \ac{DPSSB} structure to simultaneously support conventional \ac{UE} and low-power \ac{IoT} synchronization.

\bibliographystyle{IEEEtran}
\bibliography{references}

\begin{thebibliography}{10}
\providecommand{\url}[1]{#1}
\csname url@samestyle\endcsname
\providecommand{\newblock}{\relax}
\providecommand{\bibinfo}[2]{#2}
\providecommand{\BIBentrySTDinterwordspacing}{\spaceskip=0pt\relax}
\providecommand{\BIBentryALTinterwordstretchfactor}{4}
\providecommand{\BIBentryALTinterwordspacing}{\spaceskip=\fontdimen2\font plus
\BIBentryALTinterwordstretchfactor\fontdimen3\font minus \fontdimen4\font\relax}
\providecommand{\BIBforeignlanguage}[2]{{%
\expandafter\ifx\csname l@#1\endcsname\relax
\typeout{** WARNING: IEEEtran.bst: No hyphenation pattern has been}%
\typeout{** loaded for the language `#1'. Using the pattern for}%
\typeout{** the default language instead.}%
\else
\language=\csname l@#1\endcsname
\fi
#2}}
\providecommand{\BIBdecl}{\relax}
\BIBdecl

\bibitem{8766143}
Z.~Zhang, Y.~Xiao, Z.~Ma, M.~Xiao, Z.~Ding, X.~Lei, G.~K. Karagiannidis, and P.~Fan, ``{6G} wireless networks: Vision, requirements, architecture, and key technologies,'' \emph{IEEE Vehicular Technology Magazine}, vol.~14, no.~3, pp. 28--41, 2019.

\bibitem{9369324}
F.~Guo, F.~R. Yu, H.~Zhang, X.~Li, H.~Ji, and V.~C.~M. Leung, ``Enabling massive {IoT} toward {6G}: A comprehensive survey,'' \emph{IEEE Internet of Things Journal}, vol.~8, no.~15, pp. 11\,891--11\,915, 2021.

\bibitem{9200376}
A.~H. Sodhro, S.~Pirbhulal, Z.~Luo, K.~Muhammad, and N.~Z. Zahid, ``Toward {6G} architecture for energy-efficient communication in {IoT}-enabled smart automation systems,'' \emph{IEEE Internet of Things Journal}, vol.~8, no.~7, pp. 5141--5148, 2021.

\bibitem{fi11010016}
L.~Oliveira, J.~J. P.~C. Rodrigues, S.~A. Kozlov, R.~A.~L. Rabêlo, and V.~H. C.~d. Albuquerque, ``{MAC} layer protocols for internet of things: A survey,'' \emph{Future Internet}, vol.~11, no.~1, 2019.

\bibitem{9501038}
M.~Xhonneux, O.~Afisiadis, D.~Bol, and J.~Louveaux, ``A low-complexity {LoRa} synchronization algorithm robust to sampling time offsets,'' \emph{IEEE Internet of Things Journal}, vol.~9, no.~5, pp. 3756--3769, 2022.

\bibitem{barellini2023time}
A.~Barellini, B.~Bracci, G.~Licitra, and A.~M. Silvi, ``Time domain analysis of {NB}-{IoT} signals,'' \emph{Applied Sciences}, vol.~13, no.~4, p. 2242, 2023.

\bibitem{8675913}
A.~Omri, M.~Shaqfeh, A.~Ali, and H.~Alnuweiri, ``Synchronization procedure in {5G} {NR} systems,'' \emph{IEEE Access}, vol.~7, pp. 41\,286--41\,295, 2019.

\bibitem{7918505}
A.~Ali and W.~Hamouda, ``On the cell search and initial synchronization for {NB}-{IoT} {LTE} systems,'' \emph{IEEE Communications Letters}, vol.~21, no.~8, pp. 1843--1846, 2017.

\bibitem{shimura2018physical}
A.~Shimura, M.~Sawahashi, S.~Nagata, and Y.~Kishiyama, ``Physical cell {ID} detection performance applying frequency diversity reception to {NPSS} and {NSSS} for {NB}-{IoT},'' in \emph{2018 24th Asia-Pacific Conference on Communications (APCC)}.\hskip 1em plus 0.5em minus 0.4em\relax IEEE, 2018, pp. 514--519.

\bibitem{kadambar2018low}
S.~Kadambar and A.~K.~R. Chavva, ``Low complexity {ML} synchronization for {3GPP} {NB}-{IoT},'' in \emph{2018 International Conference on Signal Processing and Communications (SPCOM)}.\hskip 1em plus 0.5em minus 0.4em\relax IEEE, 2018, pp. 307--311.

\bibitem{sun2019enabling}
R.~Sun, S.~Talarico, W.~Chang, H.~Niu, and H.~Yang, ``Enabling {NB}-{IoT} on unlicensed spectrum,'' in \emph{2019 IEEE 30th Annual International Symposium on Personal, Indoor and Mobile Radio Communications (PIMRC)}.\hskip 1em plus 0.5em minus 0.4em\relax IEEE, 2019, pp. 1--7.

\bibitem{9312170}
D.~Wang, Z.~Mei, H.~Zhang, and H.~Li, ``A novel {PSS} timing synchronization algorithm for cell search in {5G} {NR} system,'' \emph{IEEE Access}, vol.~9, pp. 5870--5880, 2021.

\bibitem{ohiwa2025physical}
T.~Ohiwa, M.~Sawahashi, S.~Nagata, and S.~Suyama, ``Physical layer cell {ID} detection probabilities using {NR} synchronization signals in the presence of carrier frequency offset and phase noise in sub-{THz} bands,'' \emph{IEICE Transactions on Communications}, 2025.

\bibitem{li2022pbch}
C.~Li, H.~Li, N.~Yang, and D.~He, ``A {PBCH} reception algorithm in {5G} broadcasting,'' in \emph{2022 IEEE International Symposium on Broadband Multimedia Systems and Broadcasting (BMSB)}.\hskip 1em plus 0.5em minus 0.4em\relax IEEE, 2022, pp. 1--6.

\bibitem{chong2021efficient}
D.~Chong, G.~Ko, B.-K. Kim, J.-H. Do, and J.~Lee, ``Efficient {PBCH} {DMRS} sequence detection for fast synchronization process of {5G} {NR} systems,'' in \emph{2021 IEEE 94th Vehicular Technology Conference (VTC2021-Fall)}.\hskip 1em plus 0.5em minus 0.4em\relax IEEE, 2021, pp. 1--5.

\bibitem{10453166}
S.~Wagner, K.~L. Trung, and R.~Knopp, ``Low-power wake-up signal design in {3GPP} release 18,'' in \emph{2023 {IEEE} Conference on Standards for Communications and Networking ({CSCN})}, 2023, pp. 222--227.

\bibitem{9013137}
A.~Sahin, X.~Wang, H.~Lou, and R.~Yang, ``Low-{PAPR} multi-channel {OOK} waveform for {IEEE} 802.11ba wake-up radio,'' in \emph{2019 {IEEE} Global Communications Conference ({GLOBECOM})}, 2019, pp. 1--6.

\bibitem{11048969}
R.-A. Pitaval and X.~Tie, ``{DFT}-s-{OFDM}-based on–off keying for low-power wake-up signal,'' \emph{IEEE Transactions on Communications}, vol.~73, no.~10, pp. 9484--9500, 2025.

\bibitem{sahin2018sequence}
A.~Sahin and R.~Yang, ``Sequence-based {OOK} for orthogonal multiplexing of wake-up radio signals and {OFDM} waveforms,'' in \emph{2018 IEEE Global Communications Conference (GLOBECOM)}.\hskip 1em plus 0.5em minus 0.4em\relax IEEE, 2018, pp. 1--6.

\bibitem{zegrar2026ofdm}
H.~M. Zegrar, S.~E. Zegrar, and H.~Arslan, ``{OFDM}-based {OOK} transceiver for threshold-free detection for {WUR} and ambient {IoT} applications,'' \emph{IEEE Wireless Communications Letters}, vol.~15, pp. 1290--1294, 2026.

\bibitem{chu1972polyphase}
D.~Chu, ``Polyphase codes with good periodic correlation properties,'' \emph{IEEE Trans. Inf. Theory}, vol.~18, no.~4, pp. 531--532, 1972.

\bibitem{1204119}
F.~Digham, M.-S. Alouini, and M.~Simon, ``On the energy detection of unknown signals over fading channels,'' in \emph{IEEE International Conference on Communications, 2003. ICC '03.}, vol.~5, 2003, pp. 3575--3579 vol.5.

\bibitem{650240}
T.~Schmidl and D.~Cox, ``Robust frequency and timing synchronization for {OFDM},'' \emph{IEEE Transactions on Communications}, vol.~45, no.~12, pp. 1613--1621, 1997.

\bibitem{599949}
J.~van~de Beek, M.~Sandell, and P.~Borjesson, ``{ML} estimation of time and frequency offset in {OFDM} systems,'' \emph{IEEE Transactions on Signal Processing}, vol.~45, no.~7, pp. 1800--1805, 1997.

\bibitem{328961}
P.~Moose, ``A technique for orthogonal frequency division multiplexing frequency offset correction,'' \emph{IEEE Transactions on Communications}, vol.~42, no.~10, pp. 2908--2914, 1994.

\bibitem{9000820}
C.~Bernier, F.~Dehmas, and N.~Deparis, ``Low complexity {LoRa} frame synchronization for ultra-low power software-defined radios,'' \emph{IEEE Transactions on Communications}, vol.~68, no.~5, pp. 3140--3152, 2020.

\bibitem{tuninato2023comprehensive}
R.~Tuninato, D.~G. Riviello, R.~Garello, B.~Melis, and R.~Fantini, ``A comprehensive study on the synchronization procedure in {5G} {NR} with {3GPP}-compliant link-level simulator,'' \emph{EURASIP Journal on Wireless Communications and Networking}, vol. 2023, no.~1, p. 111, 2023.

\bibitem{1033876}
S.~Coleri, M.~Ergen, A.~Puri, and A.~Bahai, ``Channel estimation techniques based on pilot arrangement in {OFDM} systems,'' \emph{IEEE Transactions on Broadcasting}, vol.~48, no.~3, pp. 223--229, 2002.

\bibitem{504981}
J.-J. van~de Beek, O.~Edfors, M.~Sandell, S.~Wilson, and P.~Borjesson, ``On channel estimation in {OFDM} systems,'' in \emph{1995 IEEE 45th Vehicular Technology Conference. Countdown to the Wireless Twenty-First Century}, vol.~2, 1995, pp. 815--819 vol.2.

\bibitem{8032502}
J.~Zhang, M.~Wang, M.~Hua, W.~Yang, and X.~You, ``Robust synchronization waveform design for massive {IoT},'' \emph{IEEE Transactions on Wireless Communications}, vol.~16, no.~11, pp. 7551--7559, 2017.

\end{thebibliography}

\end{document}